\documentclass[journal,10pt]{IEEEtran}

\usepackage{cite}
\usepackage{color}
\usepackage{graphicx}
\usepackage{booktabs}
\usepackage{multirow}
\usepackage{subfigure}
\usepackage{comment}
\usepackage{color}
\usepackage{url}
\usepackage{amssymb}
\usepackage{amsmath}

\newcommand{\ti}{$\times$}

\begin{document}

\title{DeepQTMT: A Deep Learning Approach for Fast QTMT-based CU Partition of Intra-mode VVC}
\author{Tianyi Li, Mai Xu, \IEEEmembership{Senior Member, IEEE}, Runzhi Tang, Ying Chen and Qunliang Xing
	
	\thanks{
		This work was supported in part by the NSFC Project under Grant 62050175, Grant 61922009, and Grant 61876013; in part by the Beijing Natural Science Foundation under Grant JQ20020; and in part by the China Scholarship Council (CSC) and the Academic Excellence Foundation of Beihang University (BUAA) for Ph.D. Students. \textit{(Corresponding author: Mai Xu.)}
		
		Tianyi Li is with the School of Electronic and Information Engineering, Beihang University, Beijing 100191, China, and also with the Department of Information Technology and Electrical Engineering, ETH Zürich, 8092 Zürich, Switzerland.
		
		Mai Xu is with the School of Electronic and Information Engineering, Beihang University, Beijing 100191, China, and also with the Hangzhou Innovation Institute, Beihang University, Hangzhou 310051, China (e-mail: MaiXu@buaa.edu.cn).
		
		Runzhi Tang and Qunliang Xing are with the School of Electronic and Information Engineering, Beihang University, Beijing 100191, China.
		
		Ying Chen is with Alibaba Group, Hangzhou 311121, China.
		
		This article has supplementary downloadable material available at https://doi.org/10.1109/TIP.2021.3083447, provided by the authors.
		
		Digital Object Identifier 10.1109/TIP.2021.3083447
	}
}
\maketitle

\IEEEpeerreviewmaketitle

\begin{abstract}
Versatile Video Coding (VVC), as the latest standard, significantly improves the coding efficiency over its predecessor standard High Efficiency Video Coding (HEVC), but at the expense of sharply increased complexity. 
In VVC, the quad-tree plus multi-type tree (QTMT) structure of the coding unit (CU) partition accounts for over 97\% of the encoding time, due to the brute-force search for recursive rate-distortion (RD) optimization. 
Instead of the brute-force QTMT search, this paper proposes a deep learning approach to predict the QTMT-based CU partition, for drastically accelerating the encoding process of intra-mode VVC. 
First, we establish a large-scale database containing sufficient CU partition patterns with diverse video content, which can facilitate the data-driven VVC complexity reduction.
Next, we propose a multi-stage exit CNN (MSE-CNN) model with an early-exit mechanism to determine the CU partition, in accord with the flexible QTMT structure at multiple stages.
Then, we design an adaptive loss function for training the MSE-CNN model, synthesizing both the uncertain number of split modes and the target on minimized RD cost. 
Finally, a multi-threshold decision scheme is developed, achieving a desirable trade-off between complexity and RD performance.
The experimental results demonstrate that our approach can reduce the encoding time of VVC by 44.65\%$\sim$66.88\% with a negligible Bj\o{}ntegaard delta bit-rate (BD-BR) of 1.322\%$\sim$3.188\%, significantly outperforming other state-of-the-art approaches.
\end{abstract}


\section{Introduction}
\label{sec:intro}
Along with the development of multimedia technology, ultra-high definition (UHD) and virtual reality (VR) video have become increasingly widespread, causing the explosive growth of visual data. 
High Efficiency Video Coding (HEVC), the current-generation standard, is gradually becoming incapable for meeting the requirements of the future video market. 
Therefore, the Joint Video Exploration Team (JVET) is developing the next-generation standard, Versatile Video Coding (VVC). 
For VVC, a variety of new coding techniques have been adopted, such as the quad-tree plus multi-type tree (QTMT) structure of coding unit (CU) partition, the position-dependent intra-prediction, the affine motion compensation prediction and so on. 
These new techniques introduced in VVC achieve large gains over HEVC in coding efficiency.  
However, the complexity of VVC is also drastically greater than that of HEVC. 
As measured with the reference software VTM \cite{VTM}, the encoding complexity of VVC at intra-mode is on average 18 times higher than that of HEVC, making VVC unsuitable for practical applications. 
In particular, the QTMT-based CU partition accounts for over 97\% of the encoding time \cite{Tissier19MMSP}.
Therefore, it is necessary to significantly reduce the complexity of VVC, while maintaining the desirable coding efficiency. 

During the past decade, numerous studies have contributed to the complexity reduction of HEVC, which is the predecessor to VVC. 
In HEVC, the CU partition consumes the largest fraction of the encoding time, and thus many approaches \cite{Leng11CMSP, Shen14TIP, Zhang15TIP, Zhu17TB,Liu16TIP, Xu18TIP} have sought to simplify the CU partition in order to reduce the complexity of HEVC. 
Similarly, the CU partition structure of VVC, which is much more flexible and computationally demanding than that of HEVC, can be simplified as studied in \cite{Yamamoto16JVET, Wang17DCC, Amestoy19ICASSP, Fu19ICME_CU, Yang19TCSVT, Jin17VCIP, Wang18ICIP, Galpin19DCC}.
These studies can be classified into two categories: heuristic approaches and data-driven approaches.
In heuristic approaches, some intermediate features of encoding, such as textural homogeneity/complexity and spatial correlation, were utilized to build statistical models for the CU partition. With these models, the redundant rate-distortion optimization (RDO) processes in the earlier quad-tree plus binary-tree (QTBT) \cite{Yamamoto16JVET, Wang17DCC, Amestoy19ICASSP} or the brand-new QTMT \cite{Fu19ICME_CU, Yang19TCSVT} structure of CU partition can be skipped.
In data-driven approaches, the CU partition can be automatically learned from sufficient data, addressing the drawback that heuristic approaches rely heavily on the handcrafted feature extraction. As a representative deep learning model, the convolutional neural network (CNN) can exploit the spatial correlation of textural content. 
For example, Jin \emph{et al.} \cite{Jin17VCIP} and Wang \emph{et al.} \cite{Wang18ICIP} utilized CNN models to determine the range of CU depth in the QTBT structure. The shortcoming of \cite{Jin17VCIP, Wang18ICIP} lies in its limited potential to reduce the encoding complexity, because various CU partition patterns may have the same CU depth.
Later, Galpin \emph{et al.} \cite{Galpin19DCC} proposed directly determining the CU partition, by predicting all possible CU boundaries in units of 4$\times$4 blocks with a deep CNN. 
However, the bottom-up decision in \cite{Galpin19DCC} leads to redundant computation of the CNN, for most cases when the split CUs do not reach the minimum CU size.
Moreover, to the best of our knowledge, no data-driven approach has been designed to date for determining the newest QTMT-based CU partition in VVC.
It is worthy to directly determine the QTMT-based CU partition of VVC in a data-driven manner, benefiting from the high prediction accuracy of deep learning.

In this paper, we propose a deep learning approach to accurately predict the CU partition, in order to reduce VVC complexity at intra-mode.
First, we establish a large-scale database for learning the QTMT-based CU partition in VVC\footnote{Available online at: \url{https://github.com/tianyili2017/CPIV}}, collected from 8,000 raw images and 204 raw video sequences at four quantization parameter (QP) values.
Analyzing the sufficient data, we find that the possible split modes of CUs depend on the stage of CU partition.
Next, we propose a multi-stage exit CNN (MSE-CNN) model to determine the CU partition at multiple stages.
Combining conditional convolution in the backbone and sub-networks in the branches, the MSE-CNN model has sufficient network capacity to learn the CU partition. 
In addition, we introduce an early-exit mechanism to drastically reduce the complexity of MSE-CNN, by skipping the prediction of redundant CUs.
Furthermore, we design an adaptive loss function for training the MSE-CNN model, synthesizing both the classification loss with an uncertain number of split modes and the target on minimized rate-distortion (RD) cost. 
Finally, a multi-threshold decision scheme is developed to achieve a desirable trade-off between complexity and RD performance.
As a result, our approach can drastically reduce the complexity of intra-mode VVC, while maintaining high RD performance.
In brief, the main contributions of this paper are summarized as follows.
\begin{itemize}
	\item We establish a large-scale database to learn the QTMT-based CU partition of intra-mode VVC, which may facilitate other data-driven VVC complexity reduction studies.
	\item We propose a deep MSE-CNN model with an early-exit mechanism to determine the CU partition at multiple stages, with little computation overhead.
	\item We design an adaptive loss function synthesizing both the variable number of split modes and the optimization on RD performance, to train our MSE-CNN model.
\end{itemize}

The rest of this paper is organized as follows.
Section II reviews the related works on complexity reduction for VVC and its predecessors. 
Section III presents the database for the QTMT-based CU partition. 
In Section IV, we propose the MSE-CNN approach for fast CU partition in VVC.
Section V shows the experimental results to verify the effectiveness of our MSE-CNN approach. Finally, Section VI concludes this paper.

\section{Related Works}
\label{sec:related-work}
During the past decade, numerous approaches have been proposed to accelerate the block partition for VVC and other video coding standards. 

\subsection{Approaches for Previous Standards}
\label{sec:previous-work}
Prior to the VVC standard, some main video coding standards include HEVC, VP9 \cite{Mukherjee13PCS}, AV1 \cite{Chen18PCS} and AVS2 \cite{He13ICIP}. In particular, the HEVC standard developed by the Joint Collaborative Team on Video Coding (JCT-VC) has been established as the international video coding standard and has become a focus of research.
The approaches for simplifying the coding tree unit (CTU) partition in HEVC can be generally classified into two categories: heuristic and data-driven approaches. 
Heuristic approaches extract intermediate features during encoding to build statistical models. With these models, the brute-force RDO search of the CTU partition can be simplified, by skipping redundant processes in CTU partition. 
Considering that the CU partition consumes the most encoding time in HEVC, most approaches \cite{Leng11CMSP,Xiong14TMM,Cho13TCSVT,Shen12PCS,Kim16BMSB, Shen14TIP, Min15TCSVT, Zhang12TB, Correa15TCSVT, Hu16BMSB, Hu16TMM, Liu16DASC, Alencar16EL, Duanmu16JESTCS, Momcilovic15ISM, Du15APSIPA, Shan17ICASSP, Shen13EJIVP, Zhang15TIP, Zhu17TB, Westland19ICIP} focus on the early decision of the CU partition.
Specifically, Shen \emph{et al.} \cite{Shen14TIP} developed a dynamic CU depth range decision approach for fast intra-prediction, taking advantage of the texture property and coding information from the neighboring CUs. Then, texture homogeneity and spatial correlation are utilized to skip the coding process for some CUs. 
Min \emph{et al.} \cite{Min15TCSVT} proposed a fast CU partition prediction approach, in which global and local edge complexity is analyzed to determine the split modes of CUs.   
In addition, the support vector machine (SVM), an effective algorithm for classification, is utilized in fast CU partition.  
For example, Shen \emph{et al.} \cite{Shen13EJIVP} proposed modeling the early termination of CU partition as a binary-classification problem, utilizing a weighted SVM. 
Zhu \emph{et al.} \cite{Zhu17TB} proposed a binary and multi-class SVM approach to predict the CU partition with an off-on-line learning mechanism.
In addition to the CU partition, other recursive processes nested in CTU can be accelerated, such as prediction unit (PU) partition \cite{Khan13ICIP, Yoo13ICCE, Correa15TCSVT}, PU mode selection \cite{Jiang12ICCE, Wang13TCSVT, Duanmu16JESTCS, Lei17TB} and transform unit (TU) partition \cite{Cui17TIP, Correa15TCSVT}.

While the heuristic approaches have contributed to complexity reduction of HEVC, they rely heavily on the handcrafted feature extraction, which is inefficient in some extent and can hardly model the correlation among multiple features. In fact, the features can be automatically learned from sufficient data, benefiting from recent success of deep learning.
The CNN, as a representative deep learning model, has been utilized to reduce the complexity of CTU partition in \cite{Liu16TIP, Liu16ISCS, Laude17PCS, Xu18TIP}.
For example, Liu \emph{et al.} \cite{Liu16TIP} proposed a CNN approach for reducing the CU and PU searching modes, called the CTU structure decision CNN (CSD-CNN), such that the encoding process can be simplified. 
Laude \emph{et al.} \cite{Laude17PCS} formulated the intra-mode prediction as a multi-classification problem, and designed a five-layer CNN to select suitable prediction modes. 
Xu \emph{et al.} \cite{Xu18TIP} proposed a deep CNN model, named the early-terminated hierarchical CNN (ETH-CNN), for predicting the structured output of CU partition. As a result, the complexity for HEVC can be significantly reduced. 
Compared with the heuristic approaches, data-driven approaches typically achieve higher prediction accuracy of CTU partition, which is beneficial for the overall complexity-RD performance.
In addition to HEVC, heuristic and data-driven approaches succeed in reducing the complexity of VP9 \cite{Paul19AOMRS, Su19ICIP}, AV1 \cite{Lin18DCC, Chiang19DCC, Kim19ICASSP} and AVS2 \cite{Li17BigMM, Xie19BigMM, Yuan18BMSB, Liu19ICSIP}, by learning the binary/ternary/quad-tree based block partition structure. 

\subsection{Approaches for VVC}
\label{sec:vvc-work}

In VVC, the new partition structure of QTBT or QTMT is introduced, further enhancing the flexibility of CU partition but giving rise to extremely higher complexity. 
Similar to that of HEVC, the complexity of VVC can also be reduced by heuristic and data-driven approaches. 
For heuristic approaches \cite{Yamamoto16JVET, Wang17DCC, Amestoy19ICASSP, Amestoy20TIP, Dong21TMM, Fu19ICME_CU, Lei19ICIP, Yang19TCSVT}, earlier ones were designed for the QTBT structure.  
Among them, Yamamoto \cite{Yamamoto16JVET} proposed accelerating the QTBT-based CU partition, by reducing the maximum binary-tree depth in non-key frames. Although this scheme can reduce part of encoding time, the frame content is ignored and thus it fails to achieve the optimal CU partition in the accelerated frames.
As a solution to such content-irrelevant scheme, the decision tree was applied to select possible CU partition patterns according to the content of CUs. 
For example, Wang \emph{et al.} \cite{Wang17DCC} proposed a fast CU partition approach, using two joint decision trees to determine whether a binary/quad/binary-tree is used for each CU.
Amestoy \emph{et al.} \cite{Amestoy19ICASSP} adopted the random forest algorithm integrating multiple decision trees, for selecting one mode from binary- and quad-tree modes. 
Later, the QTMT structure has been introduced to VVC, and the corresponding approaches have also emerged.
Specifically, Fu \emph{et al.} \cite{Fu19ICME_CU} proposed a fast encoding approach, which first checks the RD cost of horizontal binary-tree mode and then decide whether to skip other split modes, based on a Bayesian-based classifier. 
Lei \emph{et al.} \cite{Lei19ICIP} developed a fast CU partition algorithm to accelerate the multi-type tree partition processes. 
Before encoding each CU, a quick RDO process is applied in advance according to the estimated intra-prediction modes with their reference pixels, named as a look-ahead mechanism, for predicting unnecessary CU partition patterns.
Yang \emph{et al.} \cite{Yang19TCSVT} proposed a fast CU partition and intra-prediction approach, through modeling the coding process as a combination of binary classifiers, based on the textural complexity of current CU and the context information from adjacent CUs. 

For data-driven approaches, Jin \emph{et al.} \cite{Jin17VCIP} utilized a CNN to predict the range of CU depth in each 32$\times$32 CU, skipping the RDO search of unused CUs at intra-mode. 
Another CNN-based approach predicting the CU depth \cite{Wang18ICIP} can be used at inter-mode, which takes a residual CU as the CNN input considering that the residue contains the correlation across adjacent frames.
Later, Galpin \emph{et al.} \cite{Galpin19DCC} proposed deciding the CU partition in a bottom-up manner, by predicting all possible CU boundaries between adjacent 4$\times$4 blocks with a deep ResNet \cite{He2016CVPR_ResNet} model. 
Different from the existing data-driven approaches \cite{Jin17VCIP, Wang18ICIP, Galpin19DCC} for VVC, we propose predicting the CU partition with a multi-stage design, providing much larger potential of complexity reduction. In addition, the proposed MSE-CNN model predicts the partition of larger CUs with former stages and that of smaller CUs with latter stages, which enables the model to early exit and avoid redundant calculation.

In this paper, we propose a deep MSE-CNN approach to predict the CU partition for intra-mode VVC, which is different from the existing deep learning approaches for HEVC/VVC performance improvement \cite{Liu16TIP, Xu18TIP, Kim19TCSVT, Kuanar18ICMEW, Kuanar19CSSP, Jin17VCIP, Wang18ICIP, Galpin19DCC, Dai17MMM, Zhang18TIP, Kuanar2018PCS} in two main aspects. 
(1) For VVC, the existing acceleration works cannot predict the QTMT-based CU partition that is finally adopted by the standard. They can only predict the range of CU depth \cite{Jin17VCIP, Wang18ICIP} or the out-of-date QTBT-based CU partition \cite{Galpin19DCC}.
(2) For HEVC, the existing acceleration works mainly focus on predicting the quad-tree-based CU partition, with either a hierarchical CU partition map \cite{Xu18TIP} or three-level classifiers deciding whether each CU is split \cite{Liu16TIP, Kim19TCSVT}. 
However, in VVC, much more splitting patterns are supported for the location and size of CUs in each CTU (5,781 patterns compared to 85 patterns), which cannot be modeled with a simple combination of classifiers. Thus, the HEVC-oriented approaches \cite{Xu18TIP, Liu16TIP, Kim19TCSVT} cannot be applied in VVC. Instead, the more complicated CU partition in VVC can be predicted by the proposed MSE-CNN model.

\section{CU Partition Database}
\label{sec:data}

\begin{figure*}[h]
	\centering  
	\includegraphics[width=.9\linewidth]{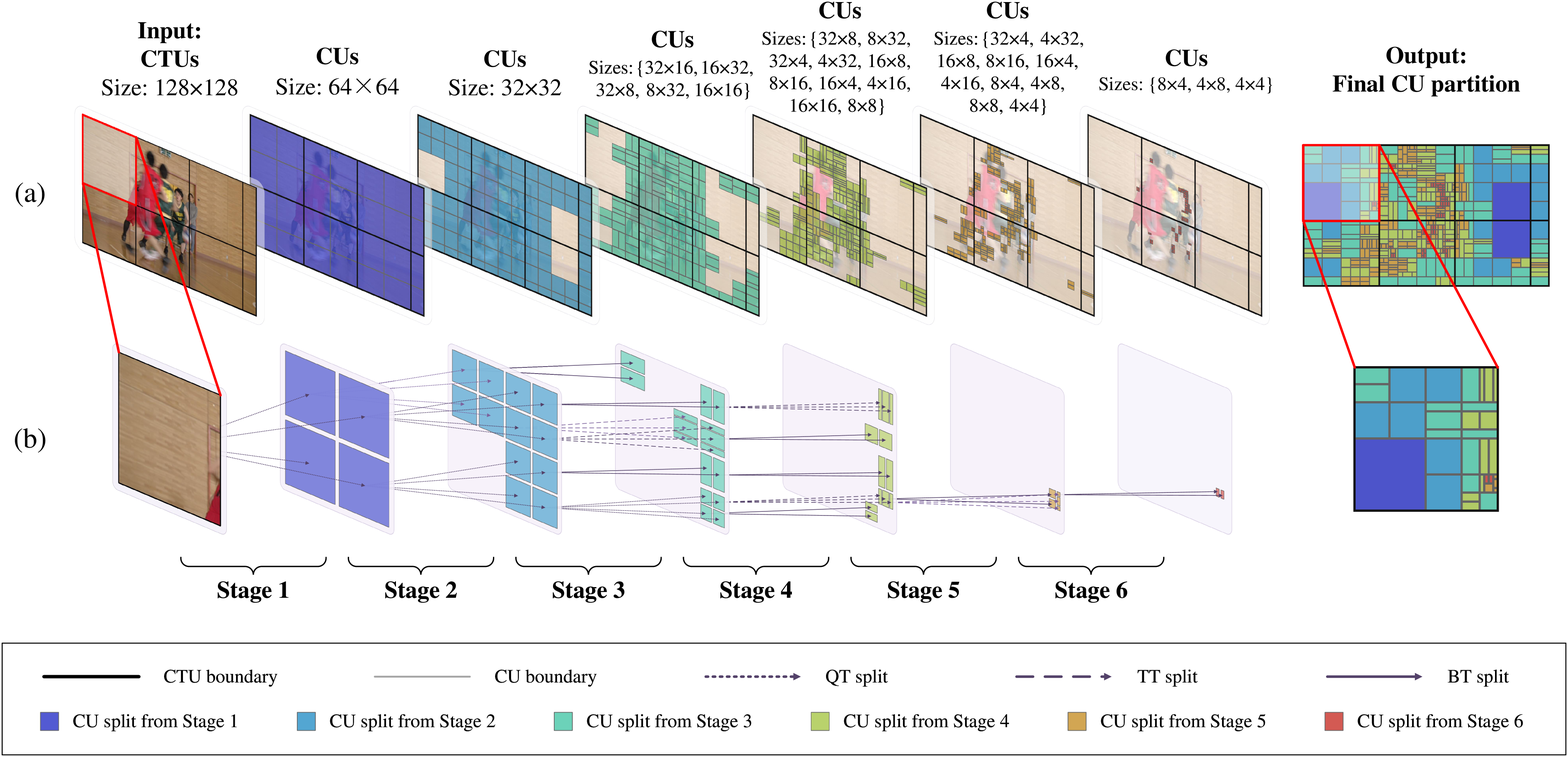}  
	\caption{Example of CU partition for luminance channel. The CUs split from different stages are distinguished by color. (a) The whole frame. (b) One CTU in this frame.} 
	\label{fig:partition_process}  
\end{figure*}

\subsection{Overview of CU Partition}
\label{sec:overview-cu}
In this section, we briefly review the CU partition in the VVC standard, which is significantly different from that in the HEVC standard.
In the HEVC standard, a CTU either contains a single CU or is recursively split into smaller square CUs via the quad-tree. 
The size of a CTU is 64$\times$64 pixels by default, and the minimal size of a CU can be 8$\times$8 in HEVC.
In the finalized VVC standard, the CU partition is more flexible than that in HEVC. With the QTMT structure, a CU can be split not only into squares, but also into rectangles.
This structure enables the CUs of VVC to be adaptive to more texture patterns of video content.
According to the QTMT structure, a CTU can either contain a single CU, or be split into smaller CUs with a quad-tree. Then, the smaller CUs can be further split with a quad-tree or multi-type tree.
The multi-type tree contains binary-tree and ternary-tree, which have two split modes, namely the horizontal and vertical modes. 
See Figure \ref{fig:partition_process} for examples. 
Additionally, the default CTU size is 128$\times$128, and the minimal size of a CU is 4$\times$4 in VVC.
Consequently, the CU sizes in the CTU are diverse, ranging from 128$\times$128 to 4$\times$4. 
Moreover, the CU partition for intra-mode VVC is separately applied for the luminance and chrominance channels, different from intra-mode HEVC where the same CU partition is used for all color channels.\footnote{In this section, we focus on the CU partition for the luminance channel, because it consumes the most encoding time in the VTM encoder \cite{VTM}. The CU partition for the chrominance channel can be analyzed in a similar manner, as shown in the \textit{Supporting Document}.}
To summarize, given the brand-new QTMT structure in VVC, the number of possible CU sizes is considerably greater than those in HEVC.

To obtain the split CUs with the above characteristics, a multi-stage hierarchical partition process is carried out.
As shown in Figure \ref{fig:partition_process}, the process of splitting a 128$\times$128 CTU into 64$\times$64 CUs can be regarded as Stage 1. Then, the process of further splitting those 64$\times$64 CUs into 32$\times$32 CUs can be regarded as Stage 2, and so on. 
Under the default configuration of intra-mode VVC, all 128$\times$128 CTUs are forced to be split into 64$\times$64 CUs, and thus only quad-tree mode is supported at Stage 1. Then, both non-splitting and quad-tree modes are supported at Stage 2.
For the subsequent stages, at most six modes are possible (non-splitting, quad-tree, horizontal binary-tree, vertical binary-tree, horizontal ternary-tree and vertical ternary-tree), satisfying that the minimum width or height is 4 for CUs.
Figure \ref{fig:partition_process} visualizes the possible CU sizes and split modes at different stages.
In the VVC standard, the optimal CU partition result is obtained through a brute-force RDO search, by checking the RD cost of all possible CUs and then selecting the combination of CUs with the minimal RD cost. The basic idea of the RDO search in VVC is similar to that in HEVC. However, the increased flexibility of CU partition in VVC leads to extremely high coding complexity compared to that for HEVC. For each CTU in HEVC, 81 CUs need to be checked during encoding, while this number increases to 5,781 in VVC. 
In fact, only a small portion of checked CUs (at least 1 CU, at most 1,024 CUs) are present in the final partition result.
Therefore, a major portion of CUs can be skipped during the RDO search, through accurate prediction of the CU partition.

\subsection{Database Establishment}
\label{sec:data-estab}

\begin{table}
	\scriptsize
	\newcommand{\tabincell}[2]{\begin{tabular}{@{}#1@{}}#2\end{tabular}}
	\begin{center}
		\caption{Configuration of CPIV database} \label{tab:database}
		\begin{tabular}{|c|c|c|c|c|}
			\hline Source & Resolution & \hspace{-0.5em}\tabincell{c}{Num. of\\ images/\\sequences}\hspace{-0.5em} & \hspace{-0.5em}\tabincell{c}{Total num.\\of CTUs}\hspace{-0.5em} & \tabincell{c}{Total num.\\of CUs}\\
			
			\hline \multirow{4}{*}{\tabincell{c}{Raw Image\\ Dataset\\ (RAISE) \cite{Dang2015MM_RAISE}}} & 2880$\times{}$1920 & 2,000 & \hspace{-0.5em}2,640,000\hspace{-0.5em} & \hspace{-0.5em}372,692,745\hspace{-0.5em} \\
			\cline{2-5} & 2304$\times{}$1536 & 2,000 & \hspace{-0.5em}1,728,000\hspace{-0.5em} & \hspace{-0.5em}242,719,640\hspace{-0.5em}  \\
			\cline{2-5} & 1536$\times{}$1024 & 2,000 & \hspace{-0.5em}768,000\hspace{-0.5em} & \hspace{-0.5em}173,216,005\hspace{-0.5em}  \\
			\cline{2-5} & 768$\times{}$512 & 2,000 & \hspace{-0.5em}192,000\hspace{-0.5em} & \hspace{-0.5em}58,271,751\hspace{-0.5em}  \\
			
			\hline \tabincell{c}{Facial video \cite{Xu2014JSTSP}} & 1920$\times$1080 (1080p) & 6 & 72,960 & \hspace{-0.5em}9,660,712\hspace{-0.5em}  \\
			
			\hline \multirow{2}{*}{\tabincell{c}{Consumer Digital\\Video Library \cite{CDVL2019}}} & 1920$\times$1080 (1080p) & 30 & \hspace{-0.5em}622,080\hspace{-0.5em} & \hspace{-0.5em}139,216,238\hspace{-0.5em} \\
			\cline{2-5} & 640$\times{}$360 (360p) & 59 & 40,520 & \hspace{-0.5em}20,699,422\hspace{-0.5em} \\
			
			\hline \multirow{7}{*}{\tabincell{c}{Xiph.org \cite{XIPH2017}}} & 2048$\times$1080 (2K) & 18 & 95,232 & 21,108,370  \\
			\cline{2-5} & 1920$\times{}$1080 (1080p) & 24 & \hspace{-0.5em}471,840\hspace{-0.5em} & \hspace{-0.5em}125,995,868\hspace{-0.5em}  \\
			\cline{2-5} & 1280$\times{}$720 (720p) & 4 & 30,600 & \hspace{-0.5em}15,913,824\hspace{-0.5em}  \\
			\cline{2-5} & 704$\times{}$576 (4CIF) & 5 & 12,400 & \hspace{-0.5em}5,411,228\hspace{-0.5em}  \\
			\cline{2-5} & 720$\times{}$486 (NTSC) & 7 & 10,545 & \hspace{-0.5em}4,765,478\hspace{-0.5em}  \\
			\cline{2-5} & 352$\times{}$288 (CIF) & 25 & 14,368 & \hspace{-0.5em}8,603,450\hspace{-0.5em}  \\
			\cline{2-5} & 352$\times{}$240 (SIF) & 4 & 688 & \hspace{-0.5em}753,882\hspace{-0.5em}  \\
			
			\hline \multicolumn{2}{|c|}{Aggregated} & 8,182 & \hspace{-0.5em}6,699,233\hspace{-0.5em} & \hspace{-0.5em}1,199,028,613\hspace{-0.5em} \\
			
			\hline
		\end{tabular}
	\end{center}
\end{table}

To train the models and evaluate the performance for our approach, we have established a large-scale database for the CU partition of intra-mode VVC (named CPIV database). The data were collected from 204 raw video sequences \cite{Xu2014JSTSP, XIPH2017, CDVL2019, Boyce18JVET} and 8,000 raw images \cite{Dang2015MM_RAISE} with multiple resolutions and diverse content. 
These video sequences and images were divided into three non-overlapping sets for training (6,400 images and 160 sequences), validation (800 images and 22 sequences) and test (800 images and 22 sequences). 
Among them, 182 training/validation sequences and all 8,000 images can be freely used for non-commercial research, and the details are listed in Table \ref{tab:database}. 
All video sequences and images were encoded by the VVC reference software VTM-7.0 \cite{VTM}. Here, four QPs $\{22, 27, 32, 37\}$ were applied to encode these sequences and images at the All-Intra (AI) configuration with the file \textit{encoder$\_$intra$\_$vtm.cfg}. 
Considering that only resolutions in multiples of 8$\times$8 are supported in VTM-7.0, the NTSC sequences were cropped to 720$\times$480 by removing the bottom edges of the frames.
Moreover, the sequences longer than 10 seconds were clipped to 10 seconds, avoiding excessively large video files in our database.

\begin{figure}[htbp]
	\centering  
	\includegraphics[width=1.0\linewidth]{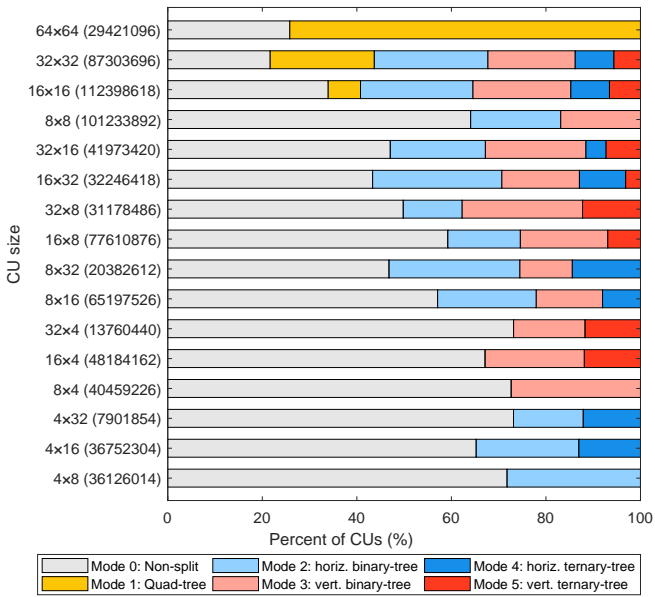}  
	\caption{Proportions of CUs with different split modes for the luminance channel. Note that each value inside parentheses represents the number of CUs. For the same CU size, the length proportion among sub-bars equals to the numerical proportion of CUs among different modes.} 
	\label{fig:cu-mode-percent}   
\end{figure}

For our CPIV database, the CU partition labels can be obtained after encoding. Each label represents the ground-truth split mode for a CU, and is equal to one of the six possible split modes: non-splitting (mode 0), quad-tree (mode 1), horizontal binary-tree (mode 2), vertical binary-tree (mode 3), horizontal ternary-tree (mode 4) and vertical ternary-tree (mode 5). 
In addition, the RD cost for all possible modes of each CU was recorded, which can be used for network training, in accord with the target of RD optimization in VVC.
Then, each CTU with the corresponding partition labels and RD cost of its CUs, forms a sample in the CPIV database.
As shown in Table \ref{tab:database}, the CPIV database contains 6,699,233 samples with more than 1 billion CUs in total, providing sufficient data for training our MSE-CNN model.
For a more detailed analysis, the proportions of CUs with different split modes\footnote{In the VTM-7.0 encoder, all 128$\times$128 CTUs are forced to be split into 64$\times$64 CUs by quad-tree. As a fixed stage, it does not need to be learned. Thus, our analysis focuses on 64$\times$64 and smaller CUs.} are illustrated in Figure \ref{fig:cu-mode-percent}.
It indicates that the number of possible modes depends on the specific CU size, ranging from 2 to 6, in agreement with the CU partition rules mentioned in Section \ref{sec:overview-cu}. Additionally, the proportions of different split modes are highly unbalanced, For example, ternary-tree split CUs (modes 4 and 5) account for less than 15\% for all CU sizes, while non-splitting CUs (mode 0) are predominant for most CU sizes. 
To solve this problem, the next section focuses on the elaborated MSE-CNN model, adaptive to the QTMT-based CU partition in the VVC standard. 

\section{Complexity Reduction for Intra-Mode VVC}
\label{sec:method}

\subsection{MSE-CNN for Learning CU Partition}
\label{sec:cnn}

\begin{figure*}
	\centering
	\subfigure[Overall structure]{
		\includegraphics[width=1\textwidth]{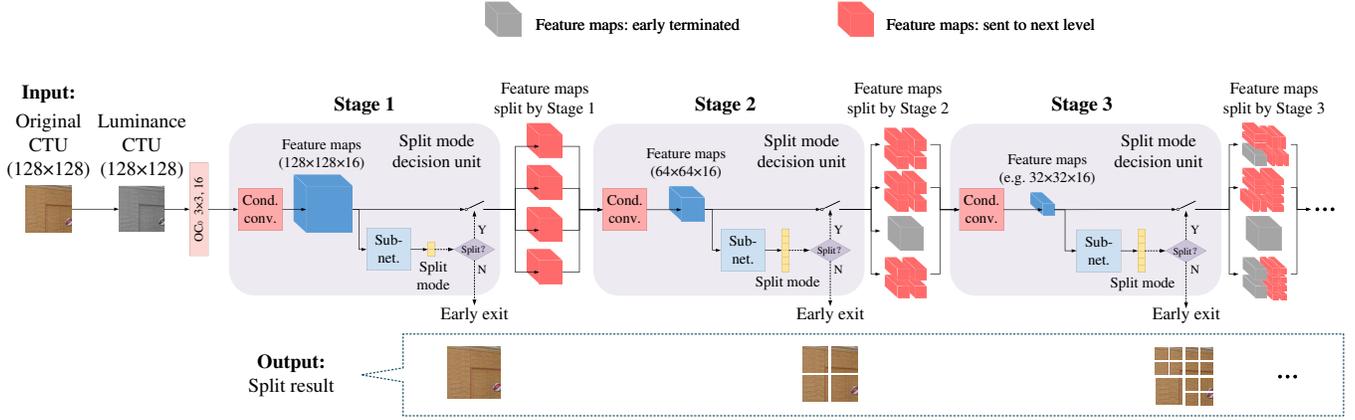}}
	\hspace{1in} 
	\subfigure[Details about conditional convolution and sub-networks]{
		\includegraphics[width=1\textwidth]{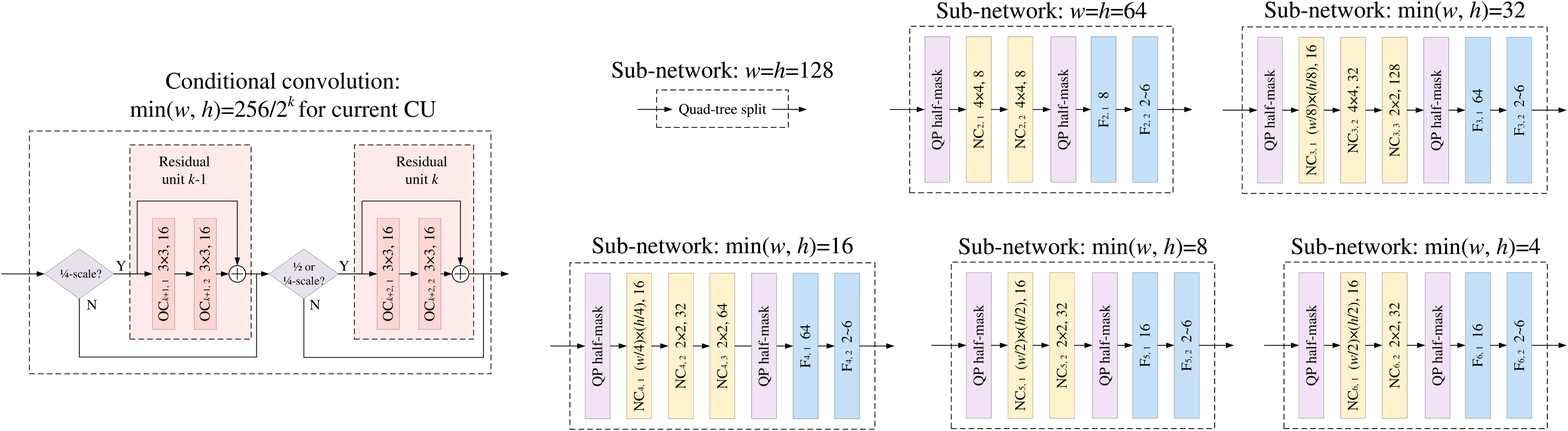}}
	\caption{Structure of MSE-CNN. The layer names started with $\mathrm{OC}$, $\mathrm{NC}$ and $\mathrm{F}$ denote overlapping convolutional, non-overlapping convolutional and fully connected layers, respectively. For convolutional layers, ``$w_\mathrm{k}\times{}h_\mathrm{k}$, $n_\mathrm{k}$'' represents $n_\mathrm{k}$ output feature maps with kernel width of $w_\mathrm{k}$ and kernel height of $h_\mathrm{k}$. For fully connected layers, the value after layer name is the number of output features.
		Note that all the convolutional and fully connected layers are activated by the parametric rectified linear units (PReLUs) \cite{He15ICCV_PReLU}, with the exception of the last fully connected layer in each sub-network activated by the Softmax function.}
	\label{fig:cnn} 
\end{figure*}

In this section, we present the proposed MSE-CNN model for learning the QTMT-based CU partition in VVC.
For the standard VVC encoder, all possible CUs in each CTU should be checked in a bottom-up manner, using the brute-force RDO search. 
In our approach, the CU partition can be predicted by MSE-CNN in a stage-wise top-down manner, to drastically accelerate the encoding process.
The overall structure of MSE-CNN is shown in Figure \ref{fig:cnn}-(a).
As shown in this figure, the luminance channel of a 128$\times$128 CTU is input to MSE-CNN, and flows through a convolutional layer to extract a group of 128$\times$128 feature maps. Using the feature maps, at most six split mode decision units are successively applied, corresponding to the CU partition at six stages. 
In each split mode decision unit, the input feature maps first flow through a series of convolutional layers, named conditional convolution, to extract the textural features in the backbone of MSE-CNN. 
Then, the feature maps are fed into a sub-network to predict the split mode of one CU, conducted in the branches of MSE-CNN. If the prediction result is non-split, the CU partition is early-terminated at the current stage; 
otherwise, the part of feature maps, corresponding to the location of each split CU, is input to the next stage.
Benefiting from the light-weighted structure of MSE-CNN, the input feature maps for all convolution operations can be directly fed into the corresponding layer, with no need to be divided into patches before convolution.
The details about conditional convolution and sub-network are presented below.

\textbf{Conditional convolution.} The efficacy of a neural network relies on sufficient features and depth. Therefore, we extract textural features and deepen the MSE-CNN model in this process. Instead of a fixed structure, we select the structure on condition of the CU size. 
This is because the CU size may be considerably variable at the same stage, and different depths of extracted features tend to be suitable for them. The mechanism of conditional convolution is shown in Figure \ref{fig:cnn}-(b), which is inspired by the efficient ResNet model \cite{He2016CVPR_ResNet}. 
Assume that the size of a CU is $w\times{}h$. Then, the minimal axis length of this CU is $\min(w,h)$, and is used to measure the granularity of the CU partition. 
If the minimal axis length of the current CU and that of its parent CU are $a_\mathrm{c}$ and $a_\mathrm{p}$, respectively, the input feature maps are processed with $n_\mathrm{r}\in\{0,1,2\}$ residual units, formulated as 
\begin{equation}
    \label{eq:resi-unit}
    n_\mathrm{r}= \left\{
	\begin{array}{ll}
		\log_{2}{(\frac{a_\mathrm{p}}{a_\mathrm{c}})} & 4 \le a_\mathrm{c} \le 64\\
		1 & a_\mathrm{c} = 128. \\
	\end{array} \right.
\end{equation} 
Here, the convolution operations in residual units are all overlapping with stride of 1 and zero-padding, keeping the size of feature maps unchanged. 
Afterwards, the feature maps processed by $n_\mathrm{r}$ residual units are used as input to the sub-network.
Such unfixed design provides a crucial property for MSE-CNN, meaning that the index of residual unit $k\in\{1,2,...,6\}$ is determinate once the size of current CU is known, satisfying $k=\log_2{[\frac{256}{\min(w,h)}]}$.  
A numerical example of conditional convolution is provided in Table \ref{tab:conv-example}. As shown in this table, the number of residual units $n_\mathrm{r}$ at each stage is decided by the minimal axis length ($a_\mathrm{p}$ and $a_\mathrm{c}$) of both the parent and current CUs, respectively. As a result, the index $k$ of residual unit ranges from 1 to 6 when $a_\mathrm{c}$ decreases from 128 to 4. For the cases other than this example, the usage of residual units can be analyzed in a similar manner.
For all residual units with the same index $k$ (though they may be at different stages) throughout MSE-CNN, we need to share the trainable parameters, ensuring that all similar-sized CUs are fed with the same sorts of features in the following sub-network. 

\begin{figure*}
	\centering
	\includegraphics[width=0.7\linewidth]{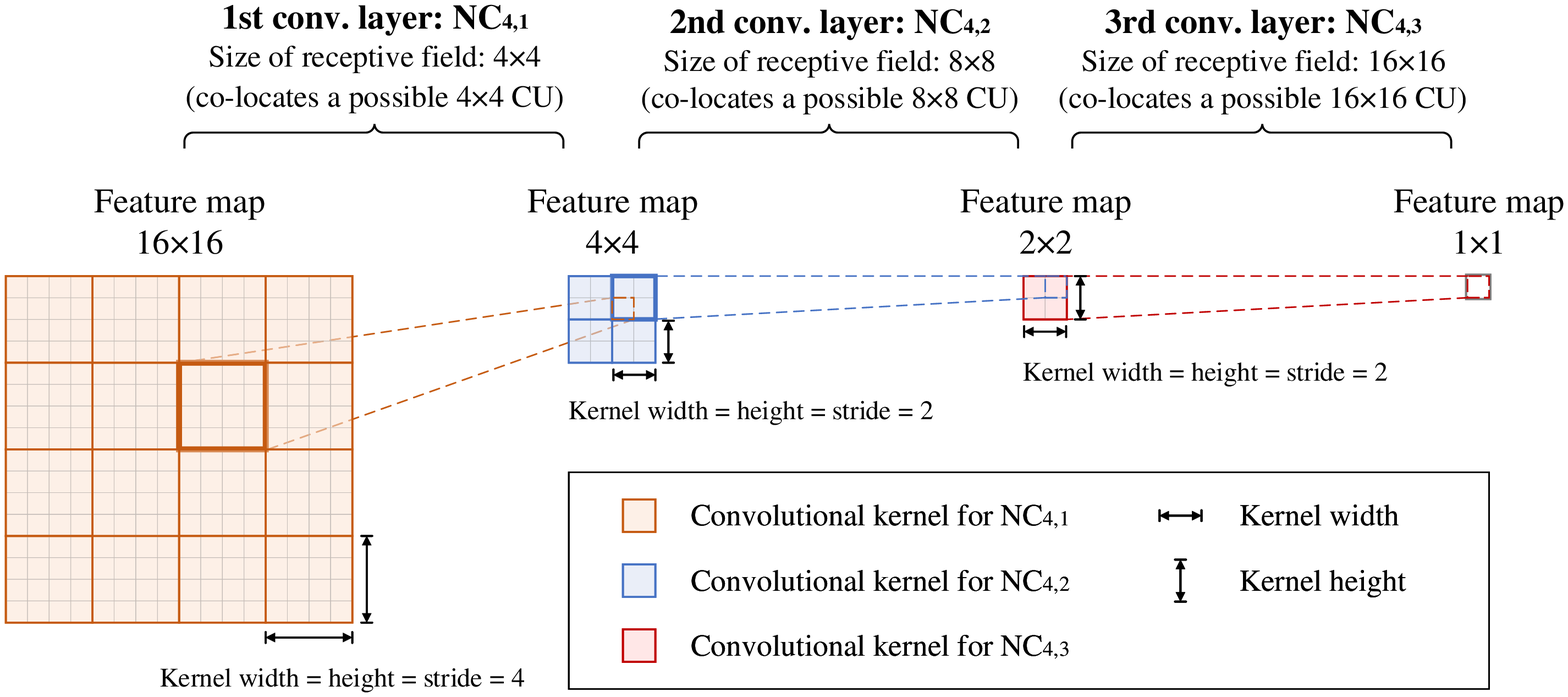}
	\caption{Non-overlapping convolution in the sub-network for partition of a 16\ti16 CU as an example. For brevity, only one feature map is shown in each group of feature maps.}
	\label{fig:cu-size-location}
\end{figure*}

\textbf{Sub-network.} In each sub-network for the partition of 64$\times$64 or smaller CUs, the input feature maps flow into a series of convolutional and fully connected layers, for predicting the split mode. The configuration of each sub-network is related to its corresponding CU size, as shown in Figure \ref{fig:cnn}-(b). 
In each sub-network, the input feature maps are fed into two or three convolutional layers, to extract low-level features for the CU partition.
For all convolutional layers, the width and height of their kernels are integer powers of 2, such as 2$\times$2 and 4$\times$4. 
Additionally, the  kernel strides in two dimensions are set equal to the width and height of the kernels, and thus all kernels are non-overlapping.
Such non-overlapping convolution is adaptive to the size and location of non-overlapping CUs in the final partition.
It lies in that the receptive field of a convolutional kernel co-locates a possible CU in most cases.  
As an example, Figure \ref{fig:cu-size-location} illustrates the convolution operations in the sub-network for partition of a 16\ti16 CU. This sub-network includes three successive non-overlapping convolutional layers $\mathrm{NC}_{4,1}$, $\mathrm{NC}_{4,2}$ and $\mathrm{NC}_{4,3}$, with the kernel sizes of 4\ti4, 2\ti2 and 2\ti2, respectively. 
The receptive fields for these layers are calculated as follow.
\begin{itemize}
	\item Direct receptive field: For the first layer $\mathrm{NC}_{4,1}$, the size of input feature map equals to that of CU, and thus the receptive field size is equal to the kernel size, 4\ti4. Because of the non-overlapping setting of convolution, these receptive fields are also non-overlapping, and the location of each receptive field is that of a possible 4\ti4 CU.
	\item Indirect receptive field: For the next layer $\mathrm{NC}_{4,2}$, considering that the input feature map has been down-sized by \ti4 scale, a 2\ti2 kernel corresponds to a receptive field of 8\ti8 in size, with each receptive field co-locating a possible 8\ti8 CU. Similarly, for layer $\mathrm{NC}_{4,3}$, the input feature map has been down-sized by \ti8 scale, and thus a 2\ti2 kernel corresponds to a 16\ti16 receptive field, co-locating a possible 16\ti16 CU.
\end{itemize}  
Moreover, an animation is provided in the \textit{Supplementary Files}\footnote{Also available online at:
	 {\url{https://github.com/tianyili2017/CPIV/blob/master/Non-overlapping_Conv.mp4}}} to better illustrate the above process.
From the above analysis, these convolutional layers are with receptive fields of 4\ti4, 8\ti8 and 16\ti16 in size, which can always be the possible location of CUs within this 16\ti16 CU. The receptive fields at each layer are all non-overlapping, and thus the CUs are non-overlapping. 
For other sub-networks with different sizes of input feature maps, there exists similar analysis.
Therefore, the non-overlapping convolution is adaptive to both the size and location of CUs.
Then, the output feature maps of convolutional layers flow through two fully connected layers to obtain the split mode. 
Consequently, each sub-network outputs a one-hot vector, which is the prediction result for the ground-truth one-hot vector.
Here, the output vector length ranges from 2 to 6, depending on the CU size. 
Moreover, QP also has a significant influence on the CU partition. Along with QP decreases, more CUs tend to be split, and vice versa. Therefore, before the first convolutional and the first fully connected layer, QP is supplemented as an external feature. Considering that some certain features in MSE-CNN may be related to QP, we apply a half-mask operation to these features, in which half of feature maps/vectors are multiplied by the normalized QP value. 
Assume that the original QP value is $q$, the normalized value is calculated by
\begin{equation}
\label{eq:qp-norm-2}
\tilde{q}=\frac{q}{51}+0.5.
\end{equation}
In the above equation, $q$ is first divided by 51, the maximum QP value in VVC, to limit its value in $[0,1]$. Then, it is added by 0.5 so that the $\tilde{q}$ is in $[0.5,1.5]$, with the average value close to 1. 
Such QP normalization is designed for ease of network training, as the numerical magnitude of feature maps/vectors almost remains the same after multiplied by $\tilde{q}$.
As such, the MSE-CNN model is able to learn the CU partition at various QP values.
Finally, the output of sub-network controls the subsequent CU partition process. If the CU is predicted as non-split, the partition process early exits at the current stage; otherwise, the output of conditional convolution at the current stage is fed into the next stage.

As discussed above, the multi-stage design combining conditional operation and sub-networks can efficiently determine the QTMT-based CU partition for the VVC standard. In addition, the early-exit mechanism drastically reduces the overall complexity of MSE-CNN, by skipping the prediction of redundant CUs. The experimental results of complexity reduction by our MSE-CNN approach are to be verified in Section \ref{sec:performance}.

\begin{table}
	\scriptsize
	\newcommand{\tabincell}[2]{\begin{tabular}{@{}#1@{}}#2\end{tabular}}
	\begin{center}
		\caption{A numerical example for conditional convolution}
		\label{tab:conv-example}
		
		\begin{tabular}{|c|c|c|c|c|c|c|}
			
			\hline Stage                                             & 1 & 2 & 3 & 4 & 5 & 6 \\
			\hline Size of current CU                                & \hspace{-0.3em}128\ti128\hspace{-0.3em} & \hspace{-0.3em}64\ti64\hspace{-0.3em}   & \hspace{-0.3em}32\ti32\hspace{-0.3em} & \hspace{-0.3em}32\ti8\hspace{-0.3em}  & \hspace{-0.3em}8\ti8\hspace{-0.3em}  & \hspace{-0.3em}4\ti8\hspace{-0.3em} \\
			\hline Size of parent CU                                 & -         & \hspace{-0.3em}128\ti128\hspace{-0.3em} & \hspace{-0.3em}64\ti64\hspace{-0.3em} & \hspace{-0.3em}32\ti32\hspace{-0.3em} & \hspace{-0.3em}32\ti8\hspace{-0.3em} & \hspace{-0.3em}8\ti8\hspace{-0.3em} \\
			\hline \tabincell{c}{Minimal axis length\\of current CU: $a_\mathrm{c}$} & 128 & 64 & 32 & 8 & 8 & 4 \\
			\hline \tabincell{c}{Minimal axis length\\of parent CU: $a_\mathrm{p}$}  & - & 128 & 64 & 32 & 8 & 8 \\
			\hline \tabincell{c}{Number of\\residual units: $n_\mathrm{r}$}                          & 1 & 1 & 1 & 2 & 0 & 1 \\
			\hline \tabincell{c}{Index of\\residual unit: $k$}                     & 1 & 2 & 3 & 4, 5 & - & 6 \\
			\hline
			
		\end{tabular}
	\end{center}
\end{table}

\subsection{Loss Function for Training MSE-CNN}
\label{sec:loss}

The proposed MSE-CNN solves a sophisticated problem with three main properties as follows. 
\begin{itemize}
	\item[(I)]  The split modes depend on the corresponding CU size, with their numbers ranging from 2 to 6.
	More details are discussed in Section \ref{sec:data}.
	\item[(II)] There exist highly unbalanced proportions for different split modes. See Figure \ref{fig:cu-mode-percent} for more details.
	\item[(III)] In VVC, different split modes typically lead to different RD cost, while a simple cross-entropy function cannot address it. 
\end{itemize}
Thus, the loss function for MSE-CNN should be adaptive to the above properties.

For a CU with width $w$ and height $h$, the set of all possible split modes is denoted by $\mathcal{M}(w,h)$. Each element $m$ in $\mathcal{M}(w,h)$ is the index of a split mode, where $m\in\{0,1,2,3,4,5\}$.
For ease of training, a mini-batch only contains CUs with the same size. Assume that the size of mini-batch is $N$, and the index of a CU is $n$.  
First, we apply the basic cross-entropy as the loss function:
\begin{equation}
\label{eq:loss-b}
L_\mathrm{CE,B}=-\frac{1}{N}\sum_{n=1}^{N}{\sum_{m\in\mathcal{M}}{y_{n,m}\log(\hat{y}_{n,m})}},
\end{equation}
where $y_{n,m}$ and $\hat{y}_{n,m}$ represent the ground-truth binary label and predicted probability for the $n$-th CU at split mode $m$.

Considering the unbalanced proportions of split modes, different penalty weights can be applied to (\ref{eq:loss-b}) according to the proportions. Then, the cross-entropy can be modified as
\begin{equation}
\label{eq:loss-w}
L_\mathrm{CE}=-\frac{\sum_{n=1}^{N}{{(\frac{1}{p_m})}^{\alpha} \cdot  \sum_{  m\in\mathcal{M}}{y_{n,m}\log(\hat{y}_{n,m})}}} {\sum_{n=1}^{N}{{(\frac{1}{p_m})}^{\alpha}}},
\end{equation}
where $p_m$ is the quantitative proportion of CUs with split mode $m$, satisfying $\sum_{m\in\mathcal{M}}{\: p_m}=1$. Additionally, $\alpha\in[0,1]$ is an adjustable scalar used to determine the importance of penalty weights. 
Here, $\alpha = 0$ means that no penalty is applied according to $\{p_m\}_{m\in\mathcal{M}}$; in this case, the MSE-CNN model may be ill-trained, because the model tends to predict only the most frequent split mode.
In contrast, $\alpha = 1$ indicates that each penalty weight is proportional to the inverse of $p_m$, avoiding the ill-trained MSE-CNN model. However, such a setting can hardly learn the prior distribution of different split modes, which may lead to a low prediction accuracy. 
As a trade-off between prediction accuracy and reliability, $\alpha\in(0,1)$ is used in practice.
In our experiments, $\alpha=0.3$ was chosen by tuning over the validation set of our CPIV database. For more details about the hyper-parameter setting, see the section of experiments.

In (\ref{eq:loss-w}), properties I and II are both addressed, while property III can be further considered by introducing a loss function of the RD cost, formulated as
\begin{equation}
\label{eq:loss-rd}
L_\mathrm{RD}=\frac{1}{N}\sum_{n=1}^{N}{\sum_{m\in\mathcal{M}}{y_{n,m}(\frac{r_{n,m}}{r_{n,\mathrm{min}}}-1)}},
\end{equation}
where $r_{n,m}$ is the RD cost for the $n$-th CU at split mode $m$, and $r_{n,\mathrm{min}}$ is the minimum RD cost for this CU among all possible split modes. 
In the above equation, $(\frac{r_{n,m}}{r_{n,\mathrm{min}}}-1)$ can be seen as the normalized RD cost. 
The term $y_{n,m}(\frac{r_{n,m}}{r_{n,\mathrm{min}}}-1)$ punishes more on either larger wrongly predicted probability $y_{n,m}$ or larger RD cost ${r_{n,m}}$, in accord with the target of RD optimization in VVC. Combining (\ref{eq:loss-w}) and (\ref{eq:loss-rd}), the overall loss function for MSE-CNN is
\begin{equation}
\label{eq:loss}
L=L_\mathrm{CE}+\beta\cdot L_\mathrm{RD}.
\end{equation}
Here, $\beta$ is a positive scalar to adjust the relative magnitude of the RD cost term over the cross-entropy term, ensuring that both terms can be effectively optimized.
As a result, the MSE-CNN model can be properly trained by minimizing $L$ of (\ref{eq:loss}).

\subsection{Multi-threshold Decision for MSE-CNN}
\label{sec:multi-thr}

Ideally, the whole CU partition is predicted by the proposed MSE-CNN model, such that all redundant checking of CUs in the original RDO process can be skipped to reduce the encoding complexity.
However, the MSE-CNN model also introduces some wrongly predicted CU partition, leading to a degradation on RD performance.
Therefore, we propose a multi-threshold decision scheme to achieve a trade-off between encoding complexity and RD performance. 

In our multi-threshold decision scheme, a combination of decision thresholds $\{\tau_s\}_{s=2}^6$, with $\tau_s$ ranging in $[0,1]$, is applied on all stages of MSE-CNN, where $s$ denotes the index of stage. 
Recall that $\hat{y}_{n,m}$ represents the predicted probability for the $n$-th CU in a mini-batch with split mode $m$, where $m$ is chosen from the possible mode set $\mathcal{M}$. In the current VTM encoder, Stage 1 is deterministic and does not need to be predicted by MSE-CNN; thus, the multi-threshold decision starts from Stage 2. 
Let the highest predicted probability be $\hat{y}_{n,\mathrm{max}}=\max \limits_{m\in\mathcal{M}}\{\hat{y}_{n,m}\}$.
For all possible modes $m\in\mathcal{M}$ of this CU, only the modes with probability $\hat{y}_{n,m}\ge\tau_s\cdot\hat{y}_{n,\mathrm{max}}$ are checked in the RDO process of the encoder, while other modes are skipped.
As such, threshold $\tau_s$ controls the confidence of MSE-CNN prediction.

\begin{figure}
	\centering  
	\includegraphics[width=1\linewidth]{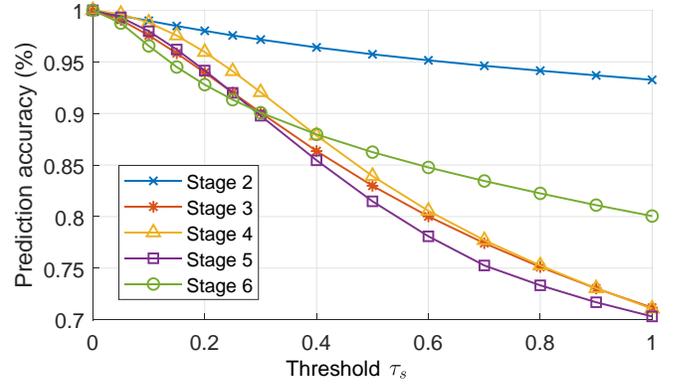}  
	\caption{Prediction accuracy of MSE-CNN on the validation data.} 
	\label{fig:accuracy-valid}   
\end{figure}

For the most aggressive setting, $\tau_s=1$ indicates that the split modes of all CUs are determined by the MSE-CNN model, as only the mode with $\hat{y}_{n,m}=\hat{y}_{n,\mathrm{max}}$ is selected for the RDO process. This setting achieves the least encoding complexity but the most degradation on RD performance.
In contrast, $\tau_s=0$ means that all CUs are checked by the original RDO process, where the encoding complexity is not reduced and there is no RD degradation.
As a trade-off, threshold $\tau_s$ is typically set between 0 and 1 in practice.
Next, we provide a scheme for selecting $\{\tau_s\}_{s=2}^6$ at the different stages of MSE-CNN, considering the unequal prediction accuracy of these stages.
Figure \ref{fig:accuracy-valid} shows the prediction accuracy of MSE-CNN with the change of $\tau_s$, averaged over all 800 images and 22 video sequences in our CPIV database (See Section V for more details about the settings). For different stages and CU sizes, the number of possible split modes may be variable (the MSE-CNN model solves a classification problem with different numbers of classes), and thus the values of Figure \ref{fig:accuracy-valid} are in the top-half accuracy.
From this figure, we can see that Stage 2 always achieves the best prediction accuracy. For Stage 6, it has the second-best prediction accuracy when the thresholds $\{\tau_s\}_{s=2}^6$ are large, while it shows relatively worse performance when the thresholds $\{\tau_s\}_{s=2}^6$ are close to 0. For other stages, the difference in accuracy is insignificant. 
Accordingly, the multi-threshold values can be chosen in the following strategies, ensuring the overall prediction accuracy of MSE-CNN. 
\begin{itemize}
\item Case 1 (more time saving): if the average threshold $\frac{1}{5}\sum\nolimits_{s=2}^6 \tau_s \ge 0.4$, then $\tau_2\ge\tau_6\ge\tau_3\approx\tau_4\approx\tau_5$.
\item Case 2 (better RD performance): if the average threshold $\frac{1}{5}\sum\nolimits_{s=2}^6 \tau_s < 0.4$, then $\tau_2\ge\tau_4\approx\tau_3\approx\tau_5\ge\tau_6$.
\end{itemize}

\section{Experimental Results}
\label{sec:result}

In this section, we conduct experiments to evaluate the effectiveness of our approach in reducing the complexity of intra-mode VVC. 
Section \ref{sec:setting} presents the experimental settings of our approach. 
Section \ref{sec:performance} evaluates the complexity and RD performance by comparing our approach with two state-of-the-art approaches \cite{Fu19ICME_CU, Yang19TCSVT}.
Then, Section \ref{sec:run-time} analyzes the complexity overhead time of our approach.
Finally, the ablation study is conducted in Section \ref{sec:ablation}.

\subsection{Configuration and Settings}
\label{sec:setting}

\textbf{Configuration of experiments.} 
In our experiments, all complexity reduction approaches were implemented in the VVC reference software VTM 7.0 \cite{VTM}.
The experiments were evaluated on all 800 test images in the CPIV database and 22 video sequences of Classes A$\sim$E in the JVET test set \cite{Boyce18JVET}. The images and sequences were encoded at the AI configuration (using the file \emph{encoder\_intra\_vtm.cfg}) at four QP values $\{22, 27, 32, 37\}$.
After encoding, $\Delta{}T$, which denotes the time-saving rate of encoding compared over the original VTM, was recorded to measure the complexity reduction. In addition, the Bj\o{}ntegaard delta bit-rate (BD-BR) and Bj\o{}ntegaard delta PSNR (BD-PSNR) \cite{Bjontegaard2001} were used to assess the RD performance.
All experiments were conducted on a computer with an Intel(R) Xeon(R) CPU E5-2650 v3 @ 2.30GHz, 128 GB RAM and the Ubuntu 18.04 64-bit operating system.
Note that a GeForce RTX 2080 Ti GPU was used to accelerate the training speed, but it was disabled when testing the encoding performance for fair comparison.

\begin{figure}
	\centering
	\includegraphics[width=1.0\linewidth]{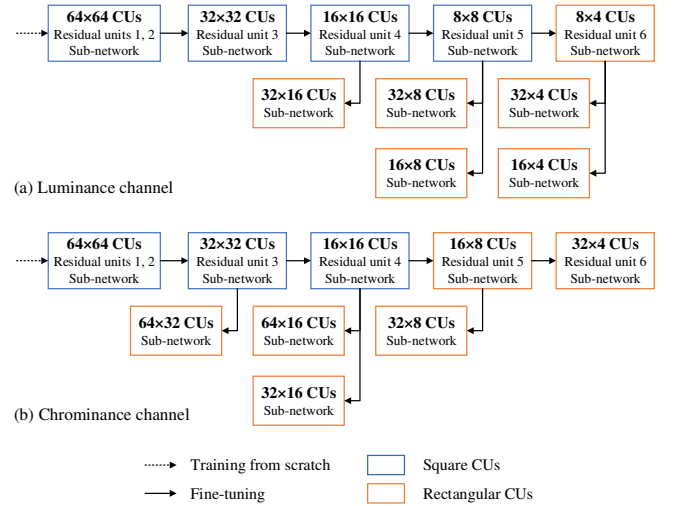}
	\caption{Training process of MES-CNN. Each block represents the training for each model.}
	\label{fig:training}
\end{figure}

\textbf{Settings for MSE-CNN.}
In intra-mode VVC, the CU partition for luminance and chrominance is determined separately by default. Thus, we trained the MSE-CNN models on different color channels separately. In total, 19 MSE-CNN models were trained according to both CU size and color channel.
Figure \ref{fig:training} illustrates the sequences of different MSE-CNN models and the trainable components in each model.
Here, all rectangular CUs fed into these models were with the width larger than the height. 
For any CU with height larger than width, it needed to be transposed in advance, in terms of both content and partition patterns.
For training MSE-CNN, all hyper-parameters were tuned on the validation set of the CPIV database.
Specifically, we set $\alpha$ and $\beta$ to be 0.3 and 1.0 in the loss function of MSE-CNN, respectively.
When training from scratch, all weight and bias parameters were randomly set with the Xavier initialization \cite{Glorot10PMLR}.
For each model trained from scratch or fine-tuning, 500,000 iterations were conducted, with the batch size of 32. 
The learning rate was initially set to $10^{-4}$ and then decreased by $1\%$ exponentially every 2,000 iterations.
During this process, the parameters in trainable components were optimized with the Adam algorithm \cite{Kingma2014CS_Adam}, while the other parameters remained unchanged.
In total, 30 hours were required to train all 19 MSE-CNN models for both luminance and chrominance channels.
For different models, the required time did not change much, always 1$\sim$2 hours for each model.
The Python 3 language and the machine learning framework PyTorch \cite{PyTorch} were used to train the MSE-CNN models, and then these models were loaded by the C++ version of PyTorch, embedded into the VTM encoder \cite{VTM} during the test.
In the inference phase of MSE-CNN, the multi-threshold values were chosen according to the analysis in Section \ref{sec:multi-thr}, as shown in Table \ref{tab:multi-thr}. Among them, the ``faster'', ``fast'' and ``medium'' modes are included.

\begin{table}[t]
	\newcommand{\tabincell}[2]{\begin{tabular}{@{}#1@{}}#2\end{tabular}}
	\begin{center}
		\caption{Multi-threshold values for MSE-CNN} \label{tab:multi-thr}
		\begin{tabular}{|c|c|c|c|c|c|c|}
			\hline \multirow{2}{*}{Mode} & \multicolumn{6}{|c|}{Threshold values} \\
			\cline{2-7} & $\tau_2$ & $\tau_3$ & $\tau_4$ & $\tau_5$ & $\tau_6$ & Average\\
			\hline ``faster''        & 0.65 & 0.45 & 0.45 & 0.45 & 0.5 & 0.5\\
			\hline ``fast'' & 0.5 & 0.4 & 0.35 & 0.35 & 0.4 & 0.4\\
			\hline ``medium''      & 0.45 & 0.3 & 0.25 & 0.25 & 0.25 & 0.3\\
			\hline
		\end{tabular}
	\end{center}
\end{table}

\subsection{Performance Evaluation}
\label{sec:performance}

In this section, we compare the performance of our MSE-CNN approach with other state-of-the-art approaches \cite{Fu19ICME_CU, Yang19TCSVT}, in both complexity reduction and coding efficiency. 
Tables \ref{tab:result-video} and \ref{tab:result-image} demonstrate the comparative results on all 22 test video sequences and 800 test images, respectively. We can see from Table \ref{tab:result-video} that the ``faster'' mode of our approach averagely reduces 59.57\%$\sim$66.88\% of encoding time on the video sequences, more effective than the time reduction of 55.65\%$\sim$59.14\% in \cite{Fu19ICME_CU} and 51.14\%$\sim$56.85\% in \cite{Yang19TCSVT}. 
For RD performance, the ``medium'' mode of our approach is with the least BD-BR increase of 1.322\% and BD-PSNR degradation of 0.055 dB on average, better than all state-of-the-art approaches \cite{Fu19ICME_CU, Yang19TCSVT}. 
Moreover, our approach in either ``faster'' or ``fast'' mode performs better than other state-of-the-art approaches. That is, ``faster'' outperforms \cite{Fu19ICME_CU} and ``fast'' outperforms \cite{Yang19TCSVT} in terms of all three metrics $\Delta T$, BD-BR and BD-PSNR. This verifies that our approach is with the best overall complexity-RD performance on video sequences. It is because the data-driven MSE-CNN model of our approach can directly predict the CU partition with high accuracy, such that most redundant processes can be skipped in the RDO search.
For images, similar results can be found in Table \ref{tab:result-image}. 

\begin{figure}
	\centering
	\includegraphics[width=1.0\linewidth]{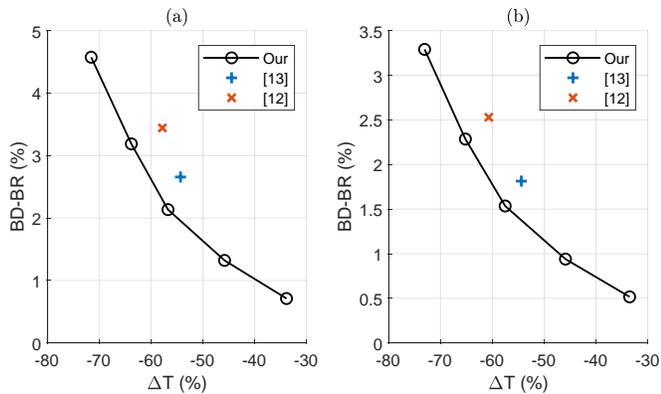}
	\caption{Complexity-RD performance for our and state-of-the-art approaches. (a) Video sequences. (b) Images.}
	\label{fig:result-curve}
\end{figure}

For a more comprehensive analysis, Figure \ref{fig:result-curve} shows the complexity-RD performance of different approaches, averaged over all four QP values.
Note that the curve of our approach is yielded by varying the multi-threshold values mentioned in Section \ref{sec:setting}. As shown in this figure, the curve of our approach locates to the bottom-left of all other approaches, for both video sequences and images. It indicates that our approach can always save more encoding time at the same BD-BR value; in other words, our approach has better RD performance with the same encoding time. Therefore, the effectiveness of our approach has been verified, and it also provides various trade-off between encoding time and RD performance.

\begin{table}
	\linespread{0.94}
	\tiny
	\newcommand{\tabincell}[2]{\begin{tabular}{@{}#1@{}}#2\end{tabular}}
	\begin{center}
		\caption{Complexity-RD performance on video sequences} \label{tab:result-video}
		\begin{tabular}{|c|c|c|c|c|c|c|c|c|}
			
			\hline \multirow{2}{*}{\hspace{-0.75em}Class\hspace{-0.75em}} & \multirow{2}{*}{Sequence} & \multirow{2}{*}{\hspace{-0.5em}Approach\hspace{-0.5em}} & \multirow{2}{*}{\tabincell{c}{\hspace{-0.5em}BD-BR\hspace{-0.5em} \\(\%)}} & \multirow{2}{*}{\tabincell{c}{\hspace{-0.75em}BD-PSNR\hspace{-0.75em} \\(dB)}} & \multicolumn{4}{|c|}{$\Delta{}T$ (\%)} \\
			\cline{6-9} & & & & & \hspace{-0.50em}QP=22\hspace{-0.50em} & \hspace{-0.50em}QP=27\hspace{-0.50em} & \hspace{-0.50em}QP=32\hspace{-0.50em} & \hspace{-0.50em}QP=37\hspace{-0.50em} \\

			  \hline \multirow{15}{*}{A1} & \multirow{5}{*}{\hspace{-0.50em}\textit{Campfire}\hspace{-0.50em}} & \cite{Fu19ICME_CU} & \hspace{-0.50em} 4.328 \hspace{-0.50em} & \hspace{-0.50em} -0.120 \hspace{-0.50em} & \hspace{-0.50em} -62.42 \hspace{-0.50em} & \hspace{-0.50em} -61.36 \hspace{-0.50em} & \hspace{-0.50em} -62.07 \hspace{-0.50em} & \hspace{-0.50em} -60.85 \hspace{-0.50em} \\
			& & \cite{Yang19TCSVT} & \hspace{-0.50em} 2.638 \hspace{-0.50em} & \hspace{-0.50em} -0.115 \hspace{-0.50em} & \hspace{-0.50em} -58.00 \hspace{-0.50em} & \hspace{-0.50em} -42.16 \hspace{-0.50em} & \hspace{-0.50em} -53.19 \hspace{-0.50em} & \hspace{-0.50em} -47.11 \hspace{-0.50em} \\
			& & \hspace{-0.75em}Our: ``faster''\hspace{-0.75em} & \hspace{-0.50em} 4.165 \hspace{-0.50em} & \hspace{-0.50em} -0.116 \hspace{-0.50em} & \hspace{-0.50em} \textbf{-65.74} \hspace{-0.50em} & \hspace{-0.50em} \textbf{-68.21} \hspace{-0.50em} & \hspace{-0.50em} \textbf{-68.02} \hspace{-0.50em} & \hspace{-0.50em} \textbf{-64.12} \hspace{-0.50em} \\
			& & \hspace{-0.75em}Our: ``fast''\hspace{-0.75em} & \hspace{-0.50em} 2.905 \hspace{-0.50em} & \hspace{-0.50em} -0.080 \hspace{-0.50em} & \hspace{-0.50em} -57.62 \hspace{-0.50em} & \hspace{-0.50em} -60.56 \hspace{-0.50em} & \hspace{-0.50em} -61.21 \hspace{-0.50em} & \hspace{-0.50em} -60.07 \hspace{-0.50em} \\
			& & \hspace{-0.75em}Our: ``medium''\hspace{-0.75em} & \hspace{-0.50em} \textbf{2.015} \hspace{-0.50em} & \hspace{-0.50em} \textbf{-0.056} \hspace{-0.50em} & \hspace{-0.50em} -43.70 \hspace{-0.50em} & \hspace{-0.50em} -47.20 \hspace{-0.50em} & \hspace{-0.50em} -52.10 \hspace{-0.50em} & \hspace{-0.50em} -51.74 \hspace{-0.50em} \\
			\cline {2-9} & \multirow{5}{*}{\hspace{-0.50em}\textit{FoodMarket4}\hspace{-0.50em}} & \cite{Fu19ICME_CU} & \hspace{-0.50em} 2.349 \hspace{-0.50em} & \hspace{-0.50em} -0.077 \hspace{-0.50em} & \hspace{-0.50em} -61.71 \hspace{-0.50em} & \hspace{-0.50em} -55.05 \hspace{-0.50em} & \hspace{-0.50em} -52.29 \hspace{-0.50em} & \hspace{-0.50em} -44.26 \hspace{-0.50em} \\
			& & \cite{Yang19TCSVT} & \hspace{-0.50em} 5.153 \hspace{-0.50em} & \hspace{-0.50em} -0.123 \hspace{-0.50em} & \hspace{-0.50em} -56.29 \hspace{-0.50em} & \hspace{-0.50em} \textbf{-74.19} \hspace{-0.50em} & \hspace{-0.50em} \textbf{-56.83} \hspace{-0.50em} & \hspace{-0.50em} \textbf{-54.63} \hspace{-0.50em} \\
			& & \hspace{-0.75em}Our: ``faster''\hspace{-0.75em} & \hspace{-0.50em} 2.784 \hspace{-0.50em} & \hspace{-0.50em} -0.091 \hspace{-0.50em} & \hspace{-0.50em} \textbf{-68.78} \hspace{-0.50em} & \hspace{-0.50em} -61.97 \hspace{-0.50em} & \hspace{-0.50em} -56.56 \hspace{-0.50em} & \hspace{-0.50em} -44.31 \hspace{-0.50em} \\
			& & \hspace{-0.75em}Our: ``fast''\hspace{-0.75em} & \hspace{-0.50em} 1.951 \hspace{-0.50em} & \hspace{-0.50em} -0.064 \hspace{-0.50em} & \hspace{-0.50em} -61.34 \hspace{-0.50em} & \hspace{-0.50em} -55.78 \hspace{-0.50em} & \hspace{-0.50em} -51.27 \hspace{-0.50em} & \hspace{-0.50em} -40.70 \hspace{-0.50em} \\
			& & \hspace{-0.75em}Our: ``medium''\hspace{-0.75em} & \hspace{-0.50em} \textbf{1.256} \hspace{-0.50em} & \hspace{-0.50em} \textbf{-0.042} \hspace{-0.50em} & \hspace{-0.50em} -49.19 \hspace{-0.50em} & \hspace{-0.50em} -44.09 \hspace{-0.50em} & \hspace{-0.50em} -42.08 \hspace{-0.50em} & \hspace{-0.50em} -33.58 \hspace{-0.50em} \\
			\cline {2-9} & \multirow{5}{*}{\hspace{-0.50em}\textit{Tango2}\hspace{-0.50em}} & \cite{Fu19ICME_CU} & \hspace{-0.50em} 3.367 \hspace{-0.50em} & \hspace{-0.50em} -0.047 \hspace{-0.50em} & \hspace{-0.50em} -65.30 \hspace{-0.50em} & \hspace{-0.50em} -59.38 \hspace{-0.50em} & \hspace{-0.50em} -49.90 \hspace{-0.50em} & \hspace{-0.50em} -35.29 \hspace{-0.50em} \\
			& & \cite{Yang19TCSVT} & \hspace{-0.50em} 1.631 \hspace{-0.50em} & \hspace{-0.50em} -0.089 \hspace{-0.50em} & \hspace{-0.50em} -51.10 \hspace{-0.50em} & \hspace{-0.50em} -37.15 \hspace{-0.50em} & \hspace{-0.50em} -48.64 \hspace{-0.50em} & \hspace{-0.50em} \textbf{-44.22} \hspace{-0.50em} \\
			& & \hspace{-0.75em}Our: ``faster''\hspace{-0.75em} & \hspace{-0.50em} 3.485 \hspace{-0.50em} & \hspace{-0.50em} -0.051 \hspace{-0.50em} & \hspace{-0.50em} \textbf{-70.83} \hspace{-0.50em} & \hspace{-0.50em} \textbf{-66.65} \hspace{-0.50em} & \hspace{-0.50em} \textbf{-53.02} \hspace{-0.50em} & \hspace{-0.50em} -26.91 \hspace{-0.50em} \\
			& & \hspace{-0.75em}Our: ``fast''\hspace{-0.75em} & \hspace{-0.50em} 2.329 \hspace{-0.50em} & \hspace{-0.50em} -0.035 \hspace{-0.50em} & \hspace{-0.50em} -64.46 \hspace{-0.50em} & \hspace{-0.50em} -60.72 \hspace{-0.50em} & \hspace{-0.50em} -47.69 \hspace{-0.50em} & \hspace{-0.50em} -25.52 \hspace{-0.50em} \\
			& & \hspace{-0.75em}Our: ``medium''\hspace{-0.75em} & \hspace{-0.50em} \textbf{1.521} \hspace{-0.50em} & \hspace{-0.50em} \textbf{-0.024} \hspace{-0.50em} & \hspace{-0.50em} -52.61 \hspace{-0.50em} & \hspace{-0.50em} -50.06 \hspace{-0.50em} & \hspace{-0.50em} -39.91 \hspace{-0.50em} & \hspace{-0.50em} -20.53 \hspace{-0.50em} \\
			
			\hline \multirow{15}{*}{A2} & \multirow{5}{*}{\hspace{-0.50em}\textit{CatRobot1}\hspace{-0.50em}} & \cite{Fu19ICME_CU} & \hspace{-0.50em} 6.748 \hspace{-0.50em} & \hspace{-0.50em} -0.152 \hspace{-0.50em} & \hspace{-0.50em} -61.81 \hspace{-0.50em} & \hspace{-0.50em} -61.95 \hspace{-0.50em} & \hspace{-0.50em} -59.75 \hspace{-0.50em} & \hspace{-0.50em} -55.49 \hspace{-0.50em} \\
			& & \cite{Yang19TCSVT} & \hspace{-0.50em} \textbf{1.128} \hspace{-0.50em} & \hspace{-0.50em} -0.066 \hspace{-0.50em} & \hspace{-0.50em} -59.21 \hspace{-0.50em} & \hspace{-0.50em} -36.92 \hspace{-0.50em} & \hspace{-0.50em} -48.00 \hspace{-0.50em} & \hspace{-0.50em} -44.55 \hspace{-0.50em} \\
			& & \hspace{-0.75em}Our: ``faster''\hspace{-0.75em} & \hspace{-0.50em} 4.875 \hspace{-0.50em} & \hspace{-0.50em} -0.112 \hspace{-0.50em} & \hspace{-0.50em} \textbf{-69.08} \hspace{-0.50em} & \hspace{-0.50em} \textbf{-64.90} \hspace{-0.50em} & \hspace{-0.50em} \textbf{-61.40} \hspace{-0.50em} & \hspace{-0.50em} \textbf{-56.11} \hspace{-0.50em} \\
			& & \hspace{-0.75em}Our: ``fast''\hspace{-0.75em} & \hspace{-0.50em} 3.282 \hspace{-0.50em} & \hspace{-0.50em} -0.078 \hspace{-0.50em} & \hspace{-0.50em} -61.29 \hspace{-0.50em} & \hspace{-0.50em} -57.10 \hspace{-0.50em} & \hspace{-0.50em} -54.06 \hspace{-0.50em} & \hspace{-0.50em} -51.51 \hspace{-0.50em} \\
			& & \hspace{-0.75em}Our: ``medium''\hspace{-0.75em} & \hspace{-0.50em} 2.163 \hspace{-0.50em} & \hspace{-0.50em} \textbf{-0.053} \hspace{-0.50em} & \hspace{-0.50em} -49.40 \hspace{-0.50em} & \hspace{-0.50em} -45.80 \hspace{-0.50em} & \hspace{-0.50em} -43.39 \hspace{-0.50em} & \hspace{-0.50em} -43.03 \hspace{-0.50em} \\
			\cline {2-9} & \multirow{5}{*}{\hspace{-0.50em}\textit{DaylightRoad2}\hspace{-0.50em}} & \cite{Fu19ICME_CU} & \hspace{-0.50em} 2.796 \hspace{-0.50em} & \hspace{-0.50em} -0.064 \hspace{-0.50em} & \hspace{-0.50em} -63.00 \hspace{-0.50em} & \hspace{-0.50em} -61.56 \hspace{-0.50em} & \hspace{-0.50em} -61.17 \hspace{-0.50em} & \hspace{-0.50em} -57.58 \hspace{-0.50em} \\
			& & \cite{Yang19TCSVT} & \hspace{-0.50em} 2.184 \hspace{-0.50em} & \hspace{-0.50em} -0.102 \hspace{-0.50em} & \hspace{-0.50em} -61.76 \hspace{-0.50em} & \hspace{-0.50em} -43.91 \hspace{-0.50em} & \hspace{-0.50em} -52.60 \hspace{-0.50em} & \hspace{-0.50em} -51.52 \hspace{-0.50em} \\
			& & \hspace{-0.75em}Our: ``faster''\hspace{-0.75em} & \hspace{-0.50em} 2.781 \hspace{-0.50em} & \hspace{-0.50em} -0.067 \hspace{-0.50em} & \hspace{-0.50em} \textbf{-72.54} \hspace{-0.50em} & \hspace{-0.50em} \textbf{-68.06} \hspace{-0.50em} & \hspace{-0.50em} \textbf{-63.65} \hspace{-0.50em} & \hspace{-0.50em} \textbf{-58.25} \hspace{-0.50em} \\
			& & \hspace{-0.75em}Our: ``fast''\hspace{-0.75em} & \hspace{-0.50em} 1.946 \hspace{-0.50em} & \hspace{-0.50em} -0.048 \hspace{-0.50em} & \hspace{-0.50em} -65.72 \hspace{-0.50em} & \hspace{-0.50em} -61.27 \hspace{-0.50em} & \hspace{-0.50em} -56.79 \hspace{-0.50em} & \hspace{-0.50em} -53.68 \hspace{-0.50em} \\
			& & \hspace{-0.75em}Our: ``medium''\hspace{-0.75em} & \hspace{-0.50em} \textbf{1.163} \hspace{-0.50em} & \hspace{-0.50em} \textbf{-0.030} \hspace{-0.50em} & \hspace{-0.50em} -56.51 \hspace{-0.50em} & \hspace{-0.50em} -49.62 \hspace{-0.50em} & \hspace{-0.50em} -45.93 \hspace{-0.50em} & \hspace{-0.50em} -45.35 \hspace{-0.50em} \\
			\cline {2-9} & \multirow{5}{*}{\hspace{-0.50em}\textit{ParkRunning3}\hspace{-0.50em}} & \cite{Fu19ICME_CU} & \hspace{-0.50em} 2.687 \hspace{-0.50em} & \hspace{-0.50em} -0.133 \hspace{-0.50em} & \hspace{-0.50em} \textbf{-64.87} \hspace{-0.50em} & \hspace{-0.50em} \textbf{-62.54} \hspace{-0.50em} & \hspace{-0.50em} \textbf{-62.40} \hspace{-0.50em} & \hspace{-0.50em} -61.80 \hspace{-0.50em} \\
			& & \cite{Yang19TCSVT} & \hspace{-0.50em} 1.247 \hspace{-0.50em} & \hspace{-0.50em} -0.081 \hspace{-0.50em} & \hspace{-0.50em} -55.38 \hspace{-0.50em} & \hspace{-0.50em} -38.99 \hspace{-0.50em} & \hspace{-0.50em} -49.76 \hspace{-0.50em} & \hspace{-0.50em} -38.17 \hspace{-0.50em} \\
			& & \hspace{-0.75em}Our: ``faster''\hspace{-0.75em} & \hspace{-0.50em} 2.675 \hspace{-0.50em} & \hspace{-0.50em} -0.132 \hspace{-0.50em} & \hspace{-0.50em} -60.50 \hspace{-0.50em} & \hspace{-0.50em} -60.47 \hspace{-0.50em} & \hspace{-0.50em} -62.09 \hspace{-0.50em} & \hspace{-0.50em} \textbf{-68.97} \hspace{-0.50em} \\
			& & \hspace{-0.75em}Our: ``fast''\hspace{-0.75em} & \hspace{-0.50em} 1.758 \hspace{-0.50em} & \hspace{-0.50em} -0.086 \hspace{-0.50em} & \hspace{-0.50em} -49.84 \hspace{-0.50em} & \hspace{-0.50em} -49.99 \hspace{-0.50em} & \hspace{-0.50em} -53.01 \hspace{-0.50em} & \hspace{-0.50em} -62.27 \hspace{-0.50em} \\
			& & \hspace{-0.75em}Our: ``medium''\hspace{-0.75em} & \hspace{-0.50em} \textbf{1.146} \hspace{-0.50em} & \hspace{-0.50em} \textbf{-0.056} \hspace{-0.50em} & \hspace{-0.50em} -35.59 \hspace{-0.50em} & \hspace{-0.50em} -36.27 \hspace{-0.50em} & \hspace{-0.50em} -41.64 \hspace{-0.50em} & \hspace{-0.50em} -53.20 \hspace{-0.50em} \\
			
			\hline \multirow{25}{*}{B} & \multirow{5}{*}{\hspace{-0.50em}\textit{MarketPlace}\hspace{-0.50em}} & \cite{Fu19ICME_CU} & \hspace{-0.50em} 2.004 \hspace{-0.50em} & \hspace{-0.50em} -0.076 \hspace{-0.50em} & \hspace{-0.50em} -61.55 \hspace{-0.50em} & \hspace{-0.50em} -60.73 \hspace{-0.50em} & \hspace{-0.50em} -59.32 \hspace{-0.50em} & \hspace{-0.50em} -58.32 \hspace{-0.50em} \\
			& & \cite{Yang19TCSVT} & \hspace{-0.50em} 4.201 \hspace{-0.50em} & \hspace{-0.50em} -0.133 \hspace{-0.50em} & \hspace{-0.50em} -48.84 \hspace{-0.50em} & \hspace{-0.50em} \textbf{-74.09} \hspace{-0.50em} & \hspace{-0.50em} -55.46 \hspace{-0.50em} & \hspace{-0.50em} -53.48 \hspace{-0.50em} \\
			& & \hspace{-0.75em}Our: ``faster''\hspace{-0.75em} & \hspace{-0.50em} 1.891 \hspace{-0.50em} & \hspace{-0.50em} -0.072 \hspace{-0.50em} & \hspace{-0.50em} \textbf{-64.84} \hspace{-0.50em} & \hspace{-0.50em} -64.42 \hspace{-0.50em} & \hspace{-0.50em} \textbf{-65.38} \hspace{-0.50em} & \hspace{-0.50em} \textbf{-65.97} \hspace{-0.50em} \\
			& & \hspace{-0.75em}Our: ``fast''\hspace{-0.75em} & \hspace{-0.50em} 1.282 \hspace{-0.50em} & \hspace{-0.50em} -0.049 \hspace{-0.50em} & \hspace{-0.50em} -55.99 \hspace{-0.50em} & \hspace{-0.50em} -56.13 \hspace{-0.50em} & \hspace{-0.50em} -58.72 \hspace{-0.50em} & \hspace{-0.50em} -62.03 \hspace{-0.50em} \\
			& & \hspace{-0.75em}Our: ``medium''\hspace{-0.75em} & \hspace{-0.50em} \textbf{0.803} \hspace{-0.50em} & \hspace{-0.50em} \textbf{-0.031} \hspace{-0.50em} & \hspace{-0.50em} -41.60 \hspace{-0.50em} & \hspace{-0.50em} -43.42 \hspace{-0.50em} & \hspace{-0.50em} -47.78 \hspace{-0.50em} & \hspace{-0.50em} -53.72 \hspace{-0.50em} \\
			\cline {2-9} & \multirow{5}{*}{\hspace{-0.50em}\textit{RitualDance}\hspace{-0.50em}} & \cite{Fu19ICME_CU} & \hspace{-0.50em} 3.859 \hspace{-0.50em} & \hspace{-0.50em} -0.183 \hspace{-0.50em} & \hspace{-0.50em} -58.74 \hspace{-0.50em} & \hspace{-0.50em} -58.32 \hspace{-0.50em} & \hspace{-0.50em} -57.22 \hspace{-0.50em} & \hspace{-0.50em} -54.40 \hspace{-0.50em} \\
			& & \cite{Yang19TCSVT} & \hspace{-0.50em} 3.731 \hspace{-0.50em} & \hspace{-0.50em} -0.113 \hspace{-0.50em} & \hspace{-0.50em} -57.72 \hspace{-0.50em} & \hspace{-0.50em} \textbf{-71.32} \hspace{-0.50em} & \hspace{-0.50em} -59.51 \hspace{-0.50em} & \hspace{-0.50em} \textbf{-59.16} \hspace{-0.50em} \\
			& & \hspace{-0.75em}Our: ``faster''\hspace{-0.75em} & \hspace{-0.50em} 2.693 \hspace{-0.50em} & \hspace{-0.50em} -0.129 \hspace{-0.50em} & \hspace{-0.50em} \textbf{-67.20} \hspace{-0.50em} & \hspace{-0.50em} -64.29 \hspace{-0.50em} & \hspace{-0.50em} \textbf{-61.78} \hspace{-0.50em} & \hspace{-0.50em} -58.64 \hspace{-0.50em} \\
			& & \hspace{-0.75em}Our: ``fast''\hspace{-0.75em} & \hspace{-0.50em} 1.796 \hspace{-0.50em} & \hspace{-0.50em} -0.086 \hspace{-0.50em} & \hspace{-0.50em} -59.06 \hspace{-0.50em} & \hspace{-0.50em} -57.74 \hspace{-0.50em} & \hspace{-0.50em} -55.67 \hspace{-0.50em} & \hspace{-0.50em} -54.54 \hspace{-0.50em} \\
			& & \hspace{-0.75em}Our: ``medium''\hspace{-0.75em} & \hspace{-0.50em} \textbf{1.071} \hspace{-0.50em} & \hspace{-0.50em} \textbf{-0.052} \hspace{-0.50em} & \hspace{-0.50em} -46.39 \hspace{-0.50em} & \hspace{-0.50em} -44.97 \hspace{-0.50em} & \hspace{-0.50em} -43.61 \hspace{-0.50em} & \hspace{-0.50em} -44.51 \hspace{-0.50em} \\
			\cline {2-9} & \multirow{5}{*}{\hspace{-0.50em}\textit{BasketballDrive}\hspace{-0.50em}} & \cite{Fu19ICME_CU} & \hspace{-0.50em} 3.553 \hspace{-0.50em} & \hspace{-0.50em} -0.091 \hspace{-0.50em} & \hspace{-0.50em} -59.71 \hspace{-0.50em} & \hspace{-0.50em} -61.18 \hspace{-0.50em} & \hspace{-0.50em} -60.12 \hspace{-0.50em} & \hspace{-0.50em} -54.79 \hspace{-0.50em} \\
			& & \cite{Yang19TCSVT} & \hspace{-0.50em} 2.185 \hspace{-0.50em} & \hspace{-0.50em} -0.055 \hspace{-0.50em} & \hspace{-0.50em} -51.80 \hspace{-0.50em} & \hspace{-0.50em} -55.68 \hspace{-0.50em} & \hspace{-0.50em} -59.96 \hspace{-0.50em} & \hspace{-0.50em} -52.82 \hspace{-0.50em} \\
			& & \hspace{-0.75em}Our: ``faster''\hspace{-0.75em} & \hspace{-0.50em} 3.873 \hspace{-0.50em} & \hspace{-0.50em} -0.101 \hspace{-0.50em} & \hspace{-0.50em} \textbf{-71.39} \hspace{-0.50em} & \hspace{-0.50em} \textbf{-72.16} \hspace{-0.50em} & \hspace{-0.50em} \textbf{-67.86} \hspace{-0.50em} & \hspace{-0.50em} \textbf{-62.60} \hspace{-0.50em} \\
			& & \hspace{-0.75em}Our: ``fast''\hspace{-0.75em} & \hspace{-0.50em} 2.613 \hspace{-0.50em} & \hspace{-0.50em} -0.069 \hspace{-0.50em} & \hspace{-0.50em} -64.48 \hspace{-0.50em} & \hspace{-0.50em} -65.93 \hspace{-0.50em} & \hspace{-0.50em} -60.62 \hspace{-0.50em} & \hspace{-0.50em} -56.95 \hspace{-0.50em} \\
			& & \hspace{-0.75em}Our: ``medium''\hspace{-0.75em} & \hspace{-0.50em} \textbf{1.642} \hspace{-0.50em} & \hspace{-0.50em} \textbf{-0.044} \hspace{-0.50em} & \hspace{-0.50em} -52.95 \hspace{-0.50em} & \hspace{-0.50em} -55.62 \hspace{-0.50em} & \hspace{-0.50em} -51.52 \hspace{-0.50em} & \hspace{-0.50em} -48.03 \hspace{-0.50em} \\
			\cline {2-9} & \multirow{5}{*}{\hspace{-0.50em}\textit{BQTerrace}\hspace{-0.50em}} & \cite{Fu19ICME_CU} & \hspace{-0.50em} 1.750 \hspace{-0.50em} & \hspace{-0.50em} -0.074 \hspace{-0.50em} & \hspace{-0.50em} -47.61 \hspace{-0.50em} & \hspace{-0.50em} -58.27 \hspace{-0.50em} & \hspace{-0.50em} -57.78 \hspace{-0.50em} & \hspace{-0.50em} -58.01 \hspace{-0.50em} \\
			& & \cite{Yang19TCSVT} & \hspace{-0.50em} 5.254 \hspace{-0.50em} & \hspace{-0.50em} -0.129 \hspace{-0.50em} & \hspace{-0.50em} -52.99 \hspace{-0.50em} & \hspace{-0.50em} -66.03 \hspace{-0.50em} & \hspace{-0.50em} -60.06 \hspace{-0.50em} & \hspace{-0.50em} -56.74 \hspace{-0.50em} \\
			& & \hspace{-0.75em}Our: ``faster''\hspace{-0.75em} & \hspace{-0.50em} 2.574 \hspace{-0.50em} & \hspace{-0.50em} -0.123 \hspace{-0.50em} & \hspace{-0.50em} \textbf{-54.21} \hspace{-0.50em} & \hspace{-0.50em} \textbf{-69.53} \hspace{-0.50em} & \hspace{-0.50em} \textbf{-67.74} \hspace{-0.50em} & \hspace{-0.50em} \textbf{-66.09} \hspace{-0.50em} \\
			& & \hspace{-0.75em}Our: ``fast''\hspace{-0.75em} & \hspace{-0.50em} 1.792 \hspace{-0.50em} & \hspace{-0.50em} -0.087 \hspace{-0.50em} & \hspace{-0.50em} -43.37 \hspace{-0.50em} & \hspace{-0.50em} -62.10 \hspace{-0.50em} & \hspace{-0.50em} -60.64 \hspace{-0.50em} & \hspace{-0.50em} -61.66 \hspace{-0.50em} \\
			& & \hspace{-0.75em}Our: ``medium''\hspace{-0.75em} & \hspace{-0.50em} \textbf{1.111} \hspace{-0.50em} & \hspace{-0.50em} \textbf{-0.054} \hspace{-0.50em} & \hspace{-0.50em} -29.37 \hspace{-0.50em} & \hspace{-0.50em} -51.23 \hspace{-0.50em} & \hspace{-0.50em} -50.41 \hspace{-0.50em} & \hspace{-0.50em} -51.45 \hspace{-0.50em} \\
			\cline {2-9} & \multirow{5}{*}{\hspace{-0.50em}\textit{Cactus}\hspace{-0.50em}} & \cite{Fu19ICME_CU} & \hspace{-0.50em} 3.541 \hspace{-0.50em} & \hspace{-0.50em} -0.112 \hspace{-0.50em} & \hspace{-0.50em} -60.10 \hspace{-0.50em} & \hspace{-0.50em} -60.11 \hspace{-0.50em} & \hspace{-0.50em} -58.78 \hspace{-0.50em} & \hspace{-0.50em} -59.22 \hspace{-0.50em} \\
			& & \cite{Yang19TCSVT} & \hspace{-0.50em} 1.315 \hspace{-0.50em} & \hspace{-0.50em} \textbf{-0.035} \hspace{-0.50em} & \hspace{-0.50em} -59.74 \hspace{-0.50em} & \hspace{-0.50em} -57.31 \hspace{-0.50em} & \hspace{-0.50em} -62.26 \hspace{-0.50em} & \hspace{-0.50em} -63.41 \hspace{-0.50em} \\
			& & \hspace{-0.75em}Our: ``faster''\hspace{-0.75em} & \hspace{-0.50em} 2.846 \hspace{-0.50em} & \hspace{-0.50em} -0.091 \hspace{-0.50em} & \hspace{-0.50em} \textbf{-69.85} \hspace{-0.50em} & \hspace{-0.50em} \textbf{-68.32} \hspace{-0.50em} & \hspace{-0.50em} \textbf{-66.23} \hspace{-0.50em} & \hspace{-0.50em} \textbf{-66.40} \hspace{-0.50em} \\
			& & \hspace{-0.75em}Our: ``fast''\hspace{-0.75em} & \hspace{-0.50em} 1.864 \hspace{-0.50em} & \hspace{-0.50em} -0.060 \hspace{-0.50em} & \hspace{-0.50em} -61.80 \hspace{-0.50em} & \hspace{-0.50em} -60.56 \hspace{-0.50em} & \hspace{-0.50em} -59.14 \hspace{-0.50em} & \hspace{-0.50em} -60.75 \hspace{-0.50em} \\
			& & \hspace{-0.75em}Our: ``medium''\hspace{-0.75em} & \hspace{-0.50em} \textbf{1.124} \hspace{-0.50em} & \hspace{-0.50em} -0.036 \hspace{-0.50em} & \hspace{-0.50em} -49.41 \hspace{-0.50em} & \hspace{-0.50em} -49.07 \hspace{-0.50em} & \hspace{-0.50em} -47.60 \hspace{-0.50em} & \hspace{-0.50em} -51.14 \hspace{-0.50em} \\
			
			\hline \multirow{20}{*}{C} & \multirow{5}{*}{\hspace{-0.50em}\textit{BasketballDrill}\hspace{-0.50em}} & \cite{Fu19ICME_CU} & \hspace{-0.50em} 4.285 \hspace{-0.50em} & \hspace{-0.50em} -0.194 \hspace{-0.50em} & \hspace{-0.50em} -56.73 \hspace{-0.50em} & \hspace{-0.50em} -60.37 \hspace{-0.50em} & \hspace{-0.50em} -59.61 \hspace{-0.50em} & \hspace{-0.50em} -56.83 \hspace{-0.50em} \\
			& & \cite{Yang19TCSVT} & \hspace{-0.50em} 4.294 \hspace{-0.50em} & \hspace{-0.50em} -0.165 \hspace{-0.50em} & \hspace{-0.50em} -60.33 \hspace{-0.50em} & \hspace{-0.50em} \textbf{-71.42} \hspace{-0.50em} & \hspace{-0.50em} -55.62 \hspace{-0.50em} & \hspace{-0.50em} -52.97 \hspace{-0.50em} \\
			& & \hspace{-0.75em}Our: ``faster''\hspace{-0.75em} & \hspace{-0.50em} 4.722 \hspace{-0.50em} & \hspace{-0.50em} -0.212 \hspace{-0.50em} & \hspace{-0.50em} \textbf{-63.15} \hspace{-0.50em} & \hspace{-0.50em} -61.55 \hspace{-0.50em} & \hspace{-0.50em} \textbf{-60.85} \hspace{-0.50em} & \hspace{-0.50em} \textbf{-58.35} \hspace{-0.50em} \\
			& & \hspace{-0.75em}Our: ``fast''\hspace{-0.75em} & \hspace{-0.50em} 2.989 \hspace{-0.50em} & \hspace{-0.50em} -0.135 \hspace{-0.50em} & \hspace{-0.50em} -53.47 \hspace{-0.50em} & \hspace{-0.50em} -53.11 \hspace{-0.50em} & \hspace{-0.50em} -53.23 \hspace{-0.50em} & \hspace{-0.50em} -50.67 \hspace{-0.50em} \\
			& & \hspace{-0.75em}Our: ``medium''\hspace{-0.75em} & \hspace{-0.50em} \textbf{1.625} \hspace{-0.50em} & \hspace{-0.50em} \textbf{-0.074} \hspace{-0.50em} & \hspace{-0.50em} -40.14 \hspace{-0.50em} & \hspace{-0.50em} -37.83 \hspace{-0.50em} & \hspace{-0.50em} -39.26 \hspace{-0.50em} & \hspace{-0.50em} -39.93 \hspace{-0.50em} \\
			\cline {2-9} & \multirow{5}{*}{\hspace{-0.50em}\textit{BQMall}\hspace{-0.50em}} & \cite{Fu19ICME_CU} & \hspace{-0.50em} 4.193 \hspace{-0.50em} & \hspace{-0.50em} -0.213 \hspace{-0.50em} & \hspace{-0.50em} -59.27 \hspace{-0.50em} & \hspace{-0.50em} -58.68 \hspace{-0.50em} & \hspace{-0.50em} -59.18 \hspace{-0.50em} & \hspace{-0.50em} -58.51 \hspace{-0.50em} \\
			& & \cite{Yang19TCSVT} & \hspace{-0.50em} 2.844 \hspace{-0.50em} & \hspace{-0.50em} -0.132 \hspace{-0.50em} & \hspace{-0.50em} -66.61 \hspace{-0.50em} & \hspace{-0.50em} -47.15 \hspace{-0.50em} & \hspace{-0.50em} -53.68 \hspace{-0.50em} & \hspace{-0.50em} -52.73 \hspace{-0.50em} \\
			& & \hspace{-0.75em}Our: ``faster''\hspace{-0.75em} & \hspace{-0.50em} 3.102 \hspace{-0.50em} & \hspace{-0.50em} -0.158 \hspace{-0.50em} & \hspace{-0.50em} \textbf{-70.67} \hspace{-0.50em} & \hspace{-0.50em} \textbf{-68.22} \hspace{-0.50em} & \hspace{-0.50em} \textbf{-65.89} \hspace{-0.50em} & \hspace{-0.50em} \textbf{-65.01} \hspace{-0.50em} \\
			& & \hspace{-0.75em}Our: ``fast''\hspace{-0.75em} & \hspace{-0.50em} 2.048 \hspace{-0.50em} & \hspace{-0.50em} -0.104 \hspace{-0.50em} & \hspace{-0.50em} -63.61 \hspace{-0.50em} & \hspace{-0.50em} -61.43 \hspace{-0.50em} & \hspace{-0.50em} -58.95 \hspace{-0.50em} & \hspace{-0.50em} -59.09 \hspace{-0.50em} \\
			& & \hspace{-0.75em}Our: ``medium''\hspace{-0.75em} & \hspace{-0.50em} \textbf{1.170} \hspace{-0.50em} & \hspace{-0.50em} \textbf{-0.060} \hspace{-0.50em} & \hspace{-0.50em} -52.86 \hspace{-0.50em} & \hspace{-0.50em} -50.51 \hspace{-0.50em} & \hspace{-0.50em} -46.87 \hspace{-0.50em} & \hspace{-0.50em} -48.78 \hspace{-0.50em} \\
			\cline {2-9} & \multirow{5}{*}{\hspace{-0.50em}\textit{PartyScene}\hspace{-0.50em}} & \cite{Fu19ICME_CU} & \hspace{-0.50em} 1.939 \hspace{-0.50em} & \hspace{-0.50em} -0.130 \hspace{-0.50em} & \hspace{-0.50em} -56.10 \hspace{-0.50em} & \hspace{-0.50em} -57.26 \hspace{-0.50em} & \hspace{-0.50em} -58.02 \hspace{-0.50em} & \hspace{-0.50em} -59.28 \hspace{-0.50em} \\
			& & \cite{Yang19TCSVT} & \hspace{-0.50em} 2.787 \hspace{-0.50em} & \hspace{-0.50em} -0.094 \hspace{-0.50em} & \hspace{-0.50em} -61.53 \hspace{-0.50em} & \hspace{-0.50em} -54.29 \hspace{-0.50em} & \hspace{-0.50em} -57.75 \hspace{-0.50em} & \hspace{-0.50em} -55.23 \hspace{-0.50em} \\
			& & \hspace{-0.75em}Our: ``faster''\hspace{-0.75em} & \hspace{-0.50em} 1.857 \hspace{-0.50em} & \hspace{-0.50em} -0.124 \hspace{-0.50em} & \hspace{-0.50em} \textbf{-65.70} \hspace{-0.50em} & \hspace{-0.50em} \textbf{-66.15} \hspace{-0.50em} & \hspace{-0.50em} \textbf{-64.20} \hspace{-0.50em} & \hspace{-0.50em} \textbf{-62.50} \hspace{-0.50em} \\
			& & \hspace{-0.75em}Our: ``fast''\hspace{-0.75em} & \hspace{-0.50em} 1.163 \hspace{-0.50em} & \hspace{-0.50em} -0.077 \hspace{-0.50em} & \hspace{-0.50em} -57.69 \hspace{-0.50em} & \hspace{-0.50em} -57.88 \hspace{-0.50em} & \hspace{-0.50em} -55.56 \hspace{-0.50em} & \hspace{-0.50em} -54.63 \hspace{-0.50em} \\
			& & \hspace{-0.75em}Our: ``medium''\hspace{-0.75em} & \hspace{-0.50em} \textbf{0.612} \hspace{-0.50em} & \hspace{-0.50em} \textbf{-0.041} \hspace{-0.50em} & \hspace{-0.50em} -47.29 \hspace{-0.50em} & \hspace{-0.50em} -47.52 \hspace{-0.50em} & \hspace{-0.50em} -43.70 \hspace{-0.50em} & \hspace{-0.50em} -42.27 \hspace{-0.50em} \\
			\cline {2-9} & \multirow{5}{*}{\hspace{-0.50em}\textit{RaceHorses}\hspace{-0.50em}} & \cite{Fu19ICME_CU} & \hspace{-0.50em} 3.181 \hspace{-0.50em} & \hspace{-0.50em} -0.171 \hspace{-0.50em} & \hspace{-0.50em} -60.73 \hspace{-0.50em} & \hspace{-0.50em} -58.76 \hspace{-0.50em} & \hspace{-0.50em} -58.39 \hspace{-0.50em} & \hspace{-0.50em} -57.71 \hspace{-0.50em} \\
			& & \cite{Yang19TCSVT} & \hspace{-0.50em} 2.394 \hspace{-0.50em} & \hspace{-0.50em} -0.097 \hspace{-0.50em} & \hspace{-0.50em} -62.38 \hspace{-0.50em} & \hspace{-0.50em} -51.08 \hspace{-0.50em} & \hspace{-0.50em} -54.89 \hspace{-0.50em} & \hspace{-0.50em} -51.29 \hspace{-0.50em} \\
			& & \hspace{-0.75em}Our: ``faster''\hspace{-0.75em} & \hspace{-0.50em} 2.503 \hspace{-0.50em} & \hspace{-0.50em} -0.135 \hspace{-0.50em} & \hspace{-0.50em} \textbf{-66.89} \hspace{-0.50em} & \hspace{-0.50em} \textbf{-65.12} \hspace{-0.50em} & \hspace{-0.50em} \textbf{-63.40} \hspace{-0.50em} & \hspace{-0.50em} \textbf{-67.31} \hspace{-0.50em} \\
			& & \hspace{-0.75em}Our: ``fast''\hspace{-0.75em} & \hspace{-0.50em} 1.611 \hspace{-0.50em} & \hspace{-0.50em} -0.086 \hspace{-0.50em} & \hspace{-0.50em} -59.01 \hspace{-0.50em} & \hspace{-0.50em} -56.21 \hspace{-0.50em} & \hspace{-0.50em} -55.25 \hspace{-0.50em} & \hspace{-0.50em} -61.09 \hspace{-0.50em} \\
			& & \hspace{-0.75em}Our: ``medium''\hspace{-0.75em} & \hspace{-0.50em} \textbf{0.963} \hspace{-0.50em} & \hspace{-0.50em} \textbf{-0.052} \hspace{-0.50em} & \hspace{-0.50em} -47.36 \hspace{-0.50em} & \hspace{-0.50em} -44.07 \hspace{-0.50em} & \hspace{-0.50em} -42.93 \hspace{-0.50em} & \hspace{-0.50em} -51.42 \hspace{-0.50em} \\
			
			\hline \multirow{20}{*}{D} & \multirow{5}{*}{\hspace{-0.50em}\textit{BasketballPass}\hspace{-0.50em}} & \cite{Fu19ICME_CU} & \hspace{-0.50em} 3.380 \hspace{-0.50em} & \hspace{-0.50em} -0.198 \hspace{-0.50em} & \hspace{-0.50em} -57.82 \hspace{-0.50em} & \hspace{-0.50em} -58.51 \hspace{-0.50em} & \hspace{-0.50em} -58.02 \hspace{-0.50em} & \hspace{-0.50em} -56.73 \hspace{-0.50em} \\
			& & \cite{Yang19TCSVT} & \hspace{-0.50em} 1.881 \hspace{-0.50em} & \hspace{-0.50em} \textbf{-0.070} \hspace{-0.50em} & \hspace{-0.50em} -51.62 \hspace{-0.50em} & \hspace{-0.50em} -51.57 \hspace{-0.50em} & \hspace{-0.50em} -62.02 \hspace{-0.50em} & \hspace{-0.50em} -53.56 \hspace{-0.50em} \\
			& & \hspace{-0.75em}Our: ``faster''\hspace{-0.75em} & \hspace{-0.50em} 3.664 \hspace{-0.50em} & \hspace{-0.50em} -0.214 \hspace{-0.50em} & \hspace{-0.50em} \textbf{-66.22} \hspace{-0.50em} & \hspace{-0.50em} \textbf{-64.96} \hspace{-0.50em} & \hspace{-0.50em} \textbf{-62.34} \hspace{-0.50em} & \hspace{-0.50em} \textbf{-56.97} \hspace{-0.50em} \\
			& & \hspace{-0.75em}Our: ``fast''\hspace{-0.75em} & \hspace{-0.50em} 2.352 \hspace{-0.50em} & \hspace{-0.50em} -0.138 \hspace{-0.50em} & \hspace{-0.50em} -58.36 \hspace{-0.50em} & \hspace{-0.50em} -57.76 \hspace{-0.50em} & \hspace{-0.50em} -55.07 \hspace{-0.50em} & \hspace{-0.50em} -49.63 \hspace{-0.50em} \\
			& & \hspace{-0.75em}Our: ``medium''\hspace{-0.75em} & \hspace{-0.50em} \textbf{1.405} \hspace{-0.50em} & \hspace{-0.50em} -0.082 \hspace{-0.50em} & \hspace{-0.50em} -47.01 \hspace{-0.50em} & \hspace{-0.50em} -46.80 \hspace{-0.50em} & \hspace{-0.50em} -44.78 \hspace{-0.50em} & \hspace{-0.50em} -39.21 \hspace{-0.50em} \\
			\cline {2-9} & \multirow{5}{*}{\hspace{-0.50em}\textit{BlowingBubbles}\hspace{-0.50em}} & \cite{Fu19ICME_CU} & \hspace{-0.50em} 2.284 \hspace{-0.50em} & \hspace{-0.50em} -0.146 \hspace{-0.50em} & \hspace{-0.50em} -54.55 \hspace{-0.50em} & \hspace{-0.50em} -54.71 \hspace{-0.50em} & \hspace{-0.50em} -54.47 \hspace{-0.50em} & \hspace{-0.50em} -57.34 \hspace{-0.50em} \\
			& & \cite{Yang19TCSVT} & \hspace{-0.50em} 3.169 \hspace{-0.50em} & \hspace{-0.50em} -0.135 \hspace{-0.50em} & \hspace{-0.50em} -50.18 \hspace{-0.50em} & \hspace{-0.50em} -61.18 \hspace{-0.50em} & \hspace{-0.50em} -52.65 \hspace{-0.50em} & \hspace{-0.50em} -45.66 \hspace{-0.50em} \\
			& & \hspace{-0.75em}Our: ``faster''\hspace{-0.75em} & \hspace{-0.50em} 2.383 \hspace{-0.50em} & \hspace{-0.50em} -0.151 \hspace{-0.50em} & \hspace{-0.50em} \textbf{-62.49} \hspace{-0.50em} & \hspace{-0.50em} \textbf{-64.10} \hspace{-0.50em} & \hspace{-0.50em} \textbf{-61.00} \hspace{-0.50em} & \hspace{-0.50em} \textbf{-59.80} \hspace{-0.50em} \\
			& & \hspace{-0.75em}Our: ``fast''\hspace{-0.75em} & \hspace{-0.50em} 1.571 \hspace{-0.50em} & \hspace{-0.50em} -0.101 \hspace{-0.50em} & \hspace{-0.50em} -52.88 \hspace{-0.50em} & \hspace{-0.50em} -55.30 \hspace{-0.50em} & \hspace{-0.50em} -52.45 \hspace{-0.50em} & \hspace{-0.50em} -52.97 \hspace{-0.50em} \\
			& & \hspace{-0.75em}Our: ``medium''\hspace{-0.75em} & \hspace{-0.50em} \textbf{0.922} \hspace{-0.50em} & \hspace{-0.50em} \textbf{-0.060} \hspace{-0.50em} & \hspace{-0.50em} -42.03 \hspace{-0.50em} & \hspace{-0.50em} -43.64 \hspace{-0.50em} & \hspace{-0.50em} -39.63 \hspace{-0.50em} & \hspace{-0.50em} -40.93 \hspace{-0.50em} \\
			\cline {2-9} & \multirow{5}{*}{\hspace{-0.50em}\textit{BQSquare}\hspace{-0.50em}} & \cite{Fu19ICME_CU} & \hspace{-0.50em} 1.134 \hspace{-0.50em} & \hspace{-0.50em} -0.083 \hspace{-0.50em} & \hspace{-0.50em} -54.63 \hspace{-0.50em} & \hspace{-0.50em} -54.41 \hspace{-0.50em} & \hspace{-0.50em} -54.10 \hspace{-0.50em} & \hspace{-0.50em} -53.54 \hspace{-0.50em} \\
			& & \cite{Yang19TCSVT} & \hspace{-0.50em} 1.365 \hspace{-0.50em} & \hspace{-0.50em} -0.085 \hspace{-0.50em} & \hspace{-0.50em} \textbf{-63.42} \hspace{-0.50em} & \hspace{-0.50em} -42.13 \hspace{-0.50em} & \hspace{-0.50em} -54.82 \hspace{-0.50em} & \hspace{-0.50em} -45.06 \hspace{-0.50em} \\
			& & \hspace{-0.75em}Our: ``faster''\hspace{-0.75em} & \hspace{-0.50em} 2.035 \hspace{-0.50em} & \hspace{-0.50em} -0.149 \hspace{-0.50em} & \hspace{-0.50em} -62.27 \hspace{-0.50em} & \hspace{-0.50em} \textbf{-61.45} \hspace{-0.50em} & \hspace{-0.50em} \textbf{-64.00} \hspace{-0.50em} & \hspace{-0.50em} \textbf{-62.36} \hspace{-0.50em} \\
			& & \hspace{-0.75em}Our: ``fast''\hspace{-0.75em} & \hspace{-0.50em} 1.327 \hspace{-0.50em} & \hspace{-0.50em} -0.097 \hspace{-0.50em} & \hspace{-0.50em} -53.25 \hspace{-0.50em} & \hspace{-0.50em} -53.89 \hspace{-0.50em} & \hspace{-0.50em} -57.45 \hspace{-0.50em} & \hspace{-0.50em} -56.05 \hspace{-0.50em} \\
			& & \hspace{-0.75em}Our: ``medium''\hspace{-0.75em} & \hspace{-0.50em} \textbf{0.743} \hspace{-0.50em} & \hspace{-0.50em} \textbf{-0.054} \hspace{-0.50em} & \hspace{-0.50em} -42.65 \hspace{-0.50em} & \hspace{-0.50em} -42.91 \hspace{-0.50em} & \hspace{-0.50em} -47.02 \hspace{-0.50em} & \hspace{-0.50em} -45.35 \hspace{-0.50em} \\
			\cline {2-9} & \multirow{5}{*}{\hspace{-0.50em}\textit{RaceHorses}\hspace{-0.50em}} & \cite{Fu19ICME_CU} & \hspace{-0.50em} 3.416 \hspace{-0.50em} & \hspace{-0.50em} -0.202 \hspace{-0.50em} & \hspace{-0.50em} -56.60 \hspace{-0.50em} & \hspace{-0.50em} -56.23 \hspace{-0.50em} & \hspace{-0.50em} -55.80 \hspace{-0.50em} & \hspace{-0.50em} -57.83 \hspace{-0.50em} \\
			& & \cite{Yang19TCSVT} & \hspace{-0.50em} \textbf{1.193} \hspace{-0.50em} & \hspace{-0.50em} \textbf{-0.066} \hspace{-0.50em} & \hspace{-0.50em} -52.86 \hspace{-0.50em} & \hspace{-0.50em} -38.00 \hspace{-0.50em} & \hspace{-0.50em} -47.55 \hspace{-0.50em} & \hspace{-0.50em} -43.45 \hspace{-0.50em} \\
			& & \hspace{-0.75em}Our: ``faster''\hspace{-0.75em} & \hspace{-0.50em} 2.917 \hspace{-0.50em} & \hspace{-0.50em} -0.171 \hspace{-0.50em} & \hspace{-0.50em} \textbf{-63.07} \hspace{-0.50em} & \hspace{-0.50em} \textbf{-60.96} \hspace{-0.50em} & \hspace{-0.50em} \textbf{-60.62} \hspace{-0.50em} & \hspace{-0.50em} \textbf{-60.51} \hspace{-0.50em} \\
			& & \hspace{-0.75em}Our: ``fast''\hspace{-0.75em} & \hspace{-0.50em} 1.880 \hspace{-0.50em} & \hspace{-0.50em} -0.110 \hspace{-0.50em} & \hspace{-0.50em} -53.50 \hspace{-0.50em} & \hspace{-0.50em} -52.70 \hspace{-0.50em} & \hspace{-0.50em} -53.01 \hspace{-0.50em} & \hspace{-0.50em} -54.01 \hspace{-0.50em} \\
			& & \hspace{-0.75em}Our: ``medium''\hspace{-0.75em} & \hspace{-0.50em} 1.200 \hspace{-0.50em} & \hspace{-0.50em} -0.071 \hspace{-0.50em} & \hspace{-0.50em} -41.59 \hspace{-0.50em} & \hspace{-0.50em} -40.72 \hspace{-0.50em} & \hspace{-0.50em} -40.76 \hspace{-0.50em} & \hspace{-0.50em} -43.49 \hspace{-0.50em} \\
			
			\hline \multirow{15}{*}{E} & \multirow{5}{*}{\hspace{-0.50em}\textit{FourPeople}\hspace{-0.50em}} & \cite{Fu19ICME_CU} & \hspace{-0.50em} 3.765 \hspace{-0.50em} & \hspace{-0.50em} -0.197 \hspace{-0.50em} & \hspace{-0.50em} -58.59 \hspace{-0.50em} & \hspace{-0.50em} -56.86 \hspace{-0.50em} & \hspace{-0.50em} -57.52 \hspace{-0.50em} & \hspace{-0.50em} -58.52 \hspace{-0.50em} \\
			& & \cite{Yang19TCSVT} & \hspace{-0.50em} 1.657 \hspace{-0.50em} & \hspace{-0.50em} -0.086 \hspace{-0.50em} & \hspace{-0.50em} -57.11 \hspace{-0.50em} & \hspace{-0.50em} -45.94 \hspace{-0.50em} & \hspace{-0.50em} -54.02 \hspace{-0.50em} & \hspace{-0.50em} -48.95 \hspace{-0.50em} \\
			& & \hspace{-0.75em}Our: ``faster''\hspace{-0.75em} & \hspace{-0.50em} 3.295 \hspace{-0.50em} & \hspace{-0.50em} -0.173 \hspace{-0.50em} & \hspace{-0.50em} \textbf{-71.63} \hspace{-0.50em} & \hspace{-0.50em} \textbf{-68.10} \hspace{-0.50em} & \hspace{-0.50em} \textbf{-64.01} \hspace{-0.50em} & \hspace{-0.50em} \textbf{-63.88} \hspace{-0.50em} \\
			& & \hspace{-0.75em}Our: ``fast''\hspace{-0.75em} & \hspace{-0.50em} 2.200 \hspace{-0.50em} & \hspace{-0.50em} -0.116 \hspace{-0.50em} & \hspace{-0.50em} -64.67 \hspace{-0.50em} & \hspace{-0.50em} -59.62 \hspace{-0.50em} & \hspace{-0.50em} -56.84 \hspace{-0.50em} & \hspace{-0.50em} -57.81 \hspace{-0.50em} \\
			& & \hspace{-0.75em}Our: ``medium''\hspace{-0.75em} & \hspace{-0.50em} \textbf{1.334} \hspace{-0.50em} & \hspace{-0.50em} \textbf{-0.070} \hspace{-0.50em} & \hspace{-0.50em} -55.33 \hspace{-0.50em} & \hspace{-0.50em} -50.24 \hspace{-0.50em} & \hspace{-0.50em} -45.80 \hspace{-0.50em} & \hspace{-0.50em} -48.12 \hspace{-0.50em} \\
			\cline {2-9} & \multirow{5}{*}{\hspace{-0.50em}\textit{Johnny}\hspace{-0.50em}} & \cite{Fu19ICME_CU} & \hspace{-0.50em} 6.479 \hspace{-0.50em} & \hspace{-0.50em} -0.240 \hspace{-0.50em} & \hspace{-0.50em} -60.76 \hspace{-0.50em} & \hspace{-0.50em} -58.49 \hspace{-0.50em} & \hspace{-0.50em} -57.56 \hspace{-0.50em} & \hspace{-0.50em} -54.35 \hspace{-0.50em} \\
			& & \cite{Yang19TCSVT} & \hspace{-0.50em} 2.428 \hspace{-0.50em} & \hspace{-0.50em} -0.110 \hspace{-0.50em} & \hspace{-0.50em} -60.33 \hspace{-0.50em} & \hspace{-0.50em} -51.29 \hspace{-0.50em} & \hspace{-0.50em} -60.38 \hspace{-0.50em} & \hspace{-0.50em} \textbf{-57.96} \hspace{-0.50em} \\
			& & \hspace{-0.75em}Our: ``faster''\hspace{-0.75em} & \hspace{-0.50em} 5.084 \hspace{-0.50em} & \hspace{-0.50em} -0.188 \hspace{-0.50em} & \hspace{-0.50em} \textbf{-71.10} \hspace{-0.50em} & \hspace{-0.50em} \textbf{-67.32} \hspace{-0.50em} & \hspace{-0.50em} \textbf{-62.69} \hspace{-0.50em} & \hspace{-0.50em} -56.29 \hspace{-0.50em} \\
			& & \hspace{-0.75em}Our: ``fast''\hspace{-0.75em} & \hspace{-0.50em} 3.565 \hspace{-0.50em} & \hspace{-0.50em} -0.132 \hspace{-0.50em} & \hspace{-0.50em} -64.04 \hspace{-0.50em} & \hspace{-0.50em} -61.21 \hspace{-0.50em} & \hspace{-0.50em} -56.75 \hspace{-0.50em} & \hspace{-0.50em} -49.53 \hspace{-0.50em} \\
			& & \hspace{-0.75em}Our: ``medium''\hspace{-0.75em} & \hspace{-0.50em} \textbf{2.327} \hspace{-0.50em} & \hspace{-0.50em} \textbf{-0.087} \hspace{-0.50em} & \hspace{-0.50em} -54.49 \hspace{-0.50em} & \hspace{-0.50em} -50.20 \hspace{-0.50em} & \hspace{-0.50em} -47.19 \hspace{-0.50em} & \hspace{-0.50em} -40.72 \hspace{-0.50em} \\
			\cline {2-9} & \multirow{5}{*}{\hspace{-0.50em}\textit{KristenAndSara}\hspace{-0.50em}} & \cite{Fu19ICME_CU} & \hspace{-0.50em} 4.707 \hspace{-0.50em} & \hspace{-0.50em} -0.215 \hspace{-0.50em} & \hspace{-0.50em} -58.47 \hspace{-0.50em} & \hspace{-0.50em} -56.60 \hspace{-0.50em} & \hspace{-0.50em} -55.88 \hspace{-0.50em} & \hspace{-0.50em} -53.71 \hspace{-0.50em} \\
			& & \cite{Yang19TCSVT} & \hspace{-0.50em} 3.782 \hspace{-0.50em} & \hspace{-0.50em} \textbf{-0.063} \hspace{-0.50em} & \hspace{-0.50em} -51.41 \hspace{-0.50em} & \hspace{-0.50em} \textbf{-69.26} \hspace{-0.50em} & \hspace{-0.50em} -62.24 \hspace{-0.50em} & \hspace{-0.50em} -52.50 \hspace{-0.50em} \\
			& & \hspace{-0.75em}Our: ``faster''\hspace{-0.75em} & \hspace{-0.50em} 3.925 \hspace{-0.50em} & \hspace{-0.50em} -0.181 \hspace{-0.50em} & \hspace{-0.50em} \textbf{-73.12} \hspace{-0.50em} & \hspace{-0.50em} -67.78 \hspace{-0.50em} & \hspace{-0.50em} \textbf{-64.36} \hspace{-0.50em} & \hspace{-0.50em} \textbf{-59.17} \hspace{-0.50em} \\
			& & \hspace{-0.75em}Our: ``fast''\hspace{-0.75em} & \hspace{-0.50em} 2.744 \hspace{-0.50em} & \hspace{-0.50em} -0.127 \hspace{-0.50em} & \hspace{-0.50em} -66.66 \hspace{-0.50em} & \hspace{-0.50em} -61.78 \hspace{-0.50em} & \hspace{-0.50em} -58.32 \hspace{-0.50em} & \hspace{-0.50em} -53.26 \hspace{-0.50em} \\
			& & \hspace{-0.75em}Our: ``medium''\hspace{-0.75em} & \hspace{-0.50em} \textbf{1.761} \hspace{-0.50em} & \hspace{-0.50em} -0.082 \hspace{-0.50em} & \hspace{-0.50em} -56.50 \hspace{-0.50em} & \hspace{-0.50em} -51.74 \hspace{-0.50em} & \hspace{-0.50em} -47.90 \hspace{-0.50em} & \hspace{-0.50em} -45.73 \hspace{-0.50em} \\
			
			\hline \multicolumn{2}{|c|}{\multirow{5}{*}{Average}} & \cite{Fu19ICME_CU} & \hspace{-0.50em} 3.443 \hspace{-0.50em} & \hspace{-0.50em} -0.142 \hspace{-0.50em} & \hspace{-0.50em} -59.14 \hspace{-0.50em} & \hspace{-0.50em} -58.70 \hspace{-0.50em} & \hspace{-0.50em} -57.70 \hspace{-0.50em} & \hspace{-0.50em} -55.65 \hspace{-0.50em} \\
			\multicolumn{2}{|c|}{} & \cite{Yang19TCSVT} & \hspace{-0.50em} 2.657 \hspace{-0.50em} & \hspace{-0.50em} -0.097 \hspace{-0.50em} & \hspace{-0.50em} -56.85 \hspace{-0.50em} & \hspace{-0.50em} -53.68 \hspace{-0.50em} & \hspace{-0.50em} -55.54 \hspace{-0.50em} & \hspace{-0.50em} -51.14 \hspace{-0.50em} \\
			\multicolumn{2}{|c|}{} & \hspace{-0.75em}Our: ``faster''\hspace{-0.75em} & \hspace{-0.50em} 3.188 \hspace{-0.50em} & \hspace{-0.50em} -0.134 \hspace{-0.50em} & \hspace{-0.50em} \textbf{-66.88} \hspace{-0.50em} & \hspace{-0.50em} \textbf{-65.67} \hspace{-0.50em} & \hspace{-0.50em} \textbf{-63.05} \hspace{-0.50em} & \hspace{-0.50em} \textbf{-59.57} \hspace{-0.50em} \\
			\multicolumn{2}{|c|}{} & \hspace{-0.75em}Our: ``fast''\hspace{-0.75em} & \hspace{-0.50em} 2.135 \hspace{-0.50em} & \hspace{-0.50em} -0.089 \hspace{-0.50em} & \hspace{-0.50em} -58.73 \hspace{-0.50em} & \hspace{-0.50em} -58.13 \hspace{-0.50em} & \hspace{-0.50em} -55.99 \hspace{-0.50em} & \hspace{-0.50em} -54.02 \hspace{-0.50em} \\
			\multicolumn{2}{|c|}{} & \hspace{-0.75em}Our: ``medium''\hspace{-0.75em} & \hspace{-0.50em} \textbf{1.322} \hspace{-0.50em} & \hspace{-0.50em} \textbf{-0.055} \hspace{-0.50em} & \hspace{-0.50em} -47.00 \hspace{-0.50em} & \hspace{-0.50em} -46.52 \hspace{-0.50em} & \hspace{-0.50em} -45.08 \hspace{-0.50em} & \hspace{-0.50em} -44.65 \hspace{-0.50em} \\
			\hline

		\end{tabular}
	\end{center}
\end{table}

\begin{table}[t]
	\linespread{0.94}
	\tiny
	\newcommand{\tabincell}[2]{\begin{tabular}{@{}#1@{}}#2\end{tabular}}
	\begin{center}
		\caption{Complexity-RD performance on images} \label{tab:result-image}
		\begin{tabular}{|c|c|c|c|c|c|c|c|c|}
			
			\hline \multirow{2}{*}{Source} & \multirow{2}{*}{Resolution} & \multirow{2}{*}{\hspace{-0.5em}Approach\hspace{-0.5em}} & \multirow{2}{*}{\tabincell{c}{\hspace{-0.5em}BD-BR\hspace{-0.5em} \\(\%)}} & \multirow{2}{*}{\tabincell{c}{\hspace{-0.75em}BD-PSNR\hspace{-0.75em} \\(dB)}} & \multicolumn{4}{|c|}{$\Delta{}T$ (\%)} \\
			\cline{6-9} & & & & & \hspace{-0.5em}QP=22\hspace{-0.5em} & \hspace{-0.5em}QP=27\hspace{-0.5em} & \hspace{-0.5em}QP=32\hspace{-0.5em} & \hspace{-0.5em}QP=37\hspace{-0.5em} \\
	
			  \hline \multirow{20}{*}{\hspace{-0.75em}\tabincell{c}{CPIV\\Database}\hspace{-0.75em}} & \multirow{5}{*}{\hspace{-0.50em}{768$\times$512}\hspace{-0.50em}} & \cite{Fu19ICME_CU} & \hspace{-0.50em} 2.115 \hspace{-0.50em} & \hspace{-0.50em} -0.104 \hspace{-0.50em} & \hspace{-0.50em} -60.89 \hspace{-0.50em} & \hspace{-0.50em} -60.45 \hspace{-0.50em} & \hspace{-0.50em} -60.16 \hspace{-0.50em} & \hspace{-0.50em} -60.33 \hspace{-0.50em} \\
			& & \cite{Yang19TCSVT} & \hspace{-0.50em} 1.226 \hspace{-0.50em} & \hspace{-0.50em} -0.065 \hspace{-0.50em} & \hspace{-0.50em} -53.47 \hspace{-0.50em} & \hspace{-0.50em} -48.38 \hspace{-0.50em} & \hspace{-0.50em} -54.18 \hspace{-0.50em} & \hspace{-0.50em} -49.92 \hspace{-0.50em} \\
			& & \hspace{-0.75em}Our: ``faster''\hspace{-0.75em} & \hspace{-0.50em} 2.234 \hspace{-0.50em} & \hspace{-0.50em} -0.111 \hspace{-0.50em} & \hspace{-0.50em} \textbf{-66.17} \hspace{-0.50em} & \hspace{-0.50em} \textbf{-64.62} \hspace{-0.50em} & \hspace{-0.50em} \textbf{-65.30} \hspace{-0.50em} & \hspace{-0.50em} \textbf{-67.46} \hspace{-0.50em} \\
			& & \hspace{-0.75em}Our: ``fast''\hspace{-0.75em} & \hspace{-0.50em} 1.469 \hspace{-0.50em} & \hspace{-0.50em} -0.073 \hspace{-0.50em} & \hspace{-0.50em} -56.67 \hspace{-0.50em} & \hspace{-0.50em} -54.96 \hspace{-0.50em} & \hspace{-0.50em} -57.21 \hspace{-0.50em} & \hspace{-0.50em} -61.17 \hspace{-0.50em} \\
			& & \hspace{-0.75em}Our: ``medium''\hspace{-0.75em} & \hspace{-0.50em} \textbf{0.838} \hspace{-0.50em} & \hspace{-0.50em} \textbf{-0.042} \hspace{-0.50em} & \hspace{-0.50em} -45.02 \hspace{-0.50em} & \hspace{-0.50em} -43.17 \hspace{-0.50em} & \hspace{-0.50em} -44.73 \hspace{-0.50em} & \hspace{-0.50em} -51.33 \hspace{-0.50em} \\
			\cline {2-9} & \multirow{5}{*}{\hspace{-0.50em}{1536$\times$1024}\hspace{-0.50em}} & \cite{Fu19ICME_CU} & \hspace{-0.50em} 2.350 \hspace{-0.50em} & \hspace{-0.50em} -0.093 \hspace{-0.50em} & \hspace{-0.50em} -61.62 \hspace{-0.50em} & \hspace{-0.50em} -60.67 \hspace{-0.50em} & \hspace{-0.50em} -60.34 \hspace{-0.50em} & \hspace{-0.50em} -59.40 \hspace{-0.50em} \\
			& & \cite{Yang19TCSVT} & \hspace{-0.50em} 1.381 \hspace{-0.50em} & \hspace{-0.50em} -0.059 \hspace{-0.50em} & \hspace{-0.50em} -54.80 \hspace{-0.50em} & \hspace{-0.50em} -51.94 \hspace{-0.50em} & \hspace{-0.50em} -57.03 \hspace{-0.50em} & \hspace{-0.50em} -52.02 \hspace{-0.50em} \\
			& & \hspace{-0.75em}Our: ``faster''\hspace{-0.75em} & \hspace{-0.50em} 2.179 \hspace{-0.50em} & \hspace{-0.50em} -0.086 \hspace{-0.50em} & \hspace{-0.50em} \textbf{-65.39} \hspace{-0.50em} & \hspace{-0.50em} \textbf{-63.54} \hspace{-0.50em} & \hspace{-0.50em} \textbf{-65.65} \hspace{-0.50em} & \hspace{-0.50em} \textbf{-67.85} \hspace{-0.50em} \\
			& & \hspace{-0.75em}Our: ``fast''\hspace{-0.75em} & \hspace{-0.50em} 1.437 \hspace{-0.50em} & \hspace{-0.50em} -0.057 \hspace{-0.50em} & \hspace{-0.50em} -56.50 \hspace{-0.50em} & \hspace{-0.50em} -55.29 \hspace{-0.50em} & \hspace{-0.50em} -58.21 \hspace{-0.50em} & \hspace{-0.50em} -62.30 \hspace{-0.50em} \\
			& & \hspace{-0.75em}Our: ``medium''\hspace{-0.75em} & \hspace{-0.50em} \textbf{0.864} \hspace{-0.50em} & \hspace{-0.50em} \textbf{-0.034} \hspace{-0.50em} & \hspace{-0.50em} -42.93 \hspace{-0.50em} & \hspace{-0.50em} -42.02 \hspace{-0.50em} & \hspace{-0.50em} -45.66 \hspace{-0.50em} & \hspace{-0.50em} -52.77 \hspace{-0.50em} \\
			\cline {2-9} & \multirow{5}{*}{\hspace{-0.50em}{2304$\times$1536}\hspace{-0.50em}} & \cite{Fu19ICME_CU} & \hspace{-0.50em} 2.787 \hspace{-0.50em} & \hspace{-0.50em} -0.121 \hspace{-0.50em} & \hspace{-0.50em} -61.45 \hspace{-0.50em} & \hspace{-0.50em} -60.71 \hspace{-0.50em} & \hspace{-0.50em} -58.72 \hspace{-0.50em} & \hspace{-0.50em} -59.05 \hspace{-0.50em} \\
			& & \cite{Yang19TCSVT} & \hspace{-0.50em} 1.721 \hspace{-0.50em} & \hspace{-0.50em} -0.079 \hspace{-0.50em} & \hspace{-0.50em} -56.23 \hspace{-0.50em} & \hspace{-0.50em} -56.19 \hspace{-0.50em} & \hspace{-0.50em} -58.01 \hspace{-0.50em} & \hspace{-0.50em} -54.29 \hspace{-0.50em} \\
			& & \hspace{-0.75em}Our: ``faster''\hspace{-0.75em} & \hspace{-0.50em} 2.369 \hspace{-0.50em} & \hspace{-0.50em} -0.103 \hspace{-0.50em} & \hspace{-0.50em} \textbf{-64.03} \hspace{-0.50em} & \hspace{-0.50em} \textbf{-63.67} \hspace{-0.50em} & \hspace{-0.50em} \textbf{-64.49} \hspace{-0.50em} & \hspace{-0.50em} \textbf{-66.02} \hspace{-0.50em} \\
			& & \hspace{-0.75em}Our: ``fast''\hspace{-0.75em} & \hspace{-0.50em} 1.622 \hspace{-0.50em} & \hspace{-0.50em} -0.071 \hspace{-0.50em} & \hspace{-0.50em} -55.36 \hspace{-0.50em} & \hspace{-0.50em} -54.92 \hspace{-0.50em} & \hspace{-0.50em} -57.79 \hspace{-0.50em} & \hspace{-0.50em} -60.70 \hspace{-0.50em} \\
			& & \hspace{-0.75em}Our: ``medium''\hspace{-0.75em} & \hspace{-0.50em} \textbf{1.022} \hspace{-0.50em} & \hspace{-0.50em} \textbf{-0.045} \hspace{-0.50em} & \hspace{-0.50em} -42.69 \hspace{-0.50em} & \hspace{-0.50em} -42.15 \hspace{-0.50em} & \hspace{-0.50em} -46.38 \hspace{-0.50em} & \hspace{-0.50em} -51.62 \hspace{-0.50em} \\
			\cline {2-9} & \multirow{5}{*}{\hspace{-0.50em}{2880$\times$1920}\hspace{-0.50em}} & \cite{Fu19ICME_CU} & \hspace{-0.50em} 2.866 \hspace{-0.50em} & \hspace{-0.50em} -0.096 \hspace{-0.50em} & \hspace{-0.50em} -64.16 \hspace{-0.50em} & \hspace{-0.50em} -63.17 \hspace{-0.50em} & \hspace{-0.50em} -61.11 \hspace{-0.50em} & \hspace{-0.50em} -58.62 \hspace{-0.50em} \\
			& & \cite{Yang19TCSVT} & \hspace{-0.50em} 2.928 \hspace{-0.50em} & \hspace{-0.50em} -0.104 \hspace{-0.50em} & \hspace{-0.50em} -51.94 \hspace{-0.50em} & \hspace{-0.50em} -56.97 \hspace{-0.50em} & \hspace{-0.50em} -59.41 \hspace{-0.50em} & \hspace{-0.50em} -55.64 \hspace{-0.50em} \\
			& & \hspace{-0.75em}Our: ``faster''\hspace{-0.75em} & \hspace{-0.50em} 2.359 \hspace{-0.50em} & \hspace{-0.50em} -0.079 \hspace{-0.50em} & \hspace{-0.50em} \textbf{-65.42} \hspace{-0.50em} & \hspace{-0.50em} \textbf{-64.35} \hspace{-0.50em} & \hspace{-0.50em} \textbf{-64.97} \hspace{-0.50em} & \hspace{-0.50em} \textbf{-64.08} \hspace{-0.50em} \\
			& & \hspace{-0.75em}Our: ``fast''\hspace{-0.75em} & \hspace{-0.50em} 1.612 \hspace{-0.50em} & \hspace{-0.50em} -0.054 \hspace{-0.50em} & \hspace{-0.50em} -55.55 \hspace{-0.50em} & \hspace{-0.50em} -56.43 \hspace{-0.50em} & \hspace{-0.50em} -58.40 \hspace{-0.50em} & \hspace{-0.50em} -58.92 \hspace{-0.50em} \\
			& & \hspace{-0.75em}Our: ``medium''\hspace{-0.75em} & \hspace{-0.50em} \textbf{1.036} \hspace{-0.50em} & \hspace{-0.50em} \textbf{-0.035} \hspace{-0.50em} & \hspace{-0.50em} -43.83 \hspace{-0.50em} & \hspace{-0.50em} -42.39 \hspace{-0.50em} & \hspace{-0.50em} -47.41 \hspace{-0.50em} & \hspace{-0.50em} -50.06 \hspace{-0.50em} \\
			
			\hline \multicolumn{2}{|c|}{\multirow{5}{*}{Average}} & \cite{Fu19ICME_CU} & \hspace{-0.50em} 2.529 \hspace{-0.50em} & \hspace{-0.50em} -0.103 \hspace{-0.50em} & \hspace{-0.50em} -62.03 \hspace{-0.50em} & \hspace{-0.50em} -61.25 \hspace{-0.50em} & \hspace{-0.50em} -60.08 \hspace{-0.50em} & \hspace{-0.50em} -59.35 \hspace{-0.50em} \\
			\multicolumn{2}{|c|}{} & \cite{Yang19TCSVT} & \hspace{-0.50em} 1.814 \hspace{-0.50em} & \hspace{-0.50em} -0.077 \hspace{-0.50em} & \hspace{-0.50em} -54.11 \hspace{-0.50em} & \hspace{-0.50em} -53.37 \hspace{-0.50em} & \hspace{-0.50em} -57.15 \hspace{-0.50em} & \hspace{-0.50em} -52.97 \hspace{-0.50em} \\
			\multicolumn{2}{|c|}{} & \hspace{-0.75em}Our: ``faster''\hspace{-0.75em} & \hspace{-0.50em} 2.285 \hspace{-0.50em} & \hspace{-0.50em} -0.095 \hspace{-0.50em} & \hspace{-0.50em} \textbf{-65.25} \hspace{-0.50em} & \hspace{-0.50em} \textbf{-64.04} \hspace{-0.50em} & \hspace{-0.50em} \textbf{-65.10} \hspace{-0.50em} & \hspace{-0.50em} \textbf{-66.35} \hspace{-0.50em} \\
			\multicolumn{2}{|c|}{} & \hspace{-0.75em}Our: ``fast''\hspace{-0.75em} & \hspace{-0.50em} 1.535 \hspace{-0.50em} & \hspace{-0.50em} -0.064 \hspace{-0.50em} & \hspace{-0.50em} -56.02 \hspace{-0.50em} & \hspace{-0.50em} -55.40 \hspace{-0.50em} & \hspace{-0.50em} -57.90 \hspace{-0.50em} & \hspace{-0.50em} -60.77 \hspace{-0.50em} \\
			\multicolumn{2}{|c|}{} & \hspace{-0.75em}Our: ``medium''\hspace{-0.75em} & \hspace{-0.50em} \textbf{0.940} \hspace{-0.50em} & \hspace{-0.50em} \textbf{-0.039} \hspace{-0.50em} & \hspace{-0.50em} -43.62 \hspace{-0.50em} & \hspace{-0.50em} -42.43 \hspace{-0.50em} & \hspace{-0.50em} -46.04 \hspace{-0.50em} & \hspace{-0.50em} -51.45 \hspace{-0.50em} \\
			\hline 
			
		\end{tabular}
	\end{center}
\end{table}

\subsection{Complexity Overhead Analysis}
\label{sec:run-time}

\begin{figure}
	\centering
	\includegraphics[width=1.0\linewidth]{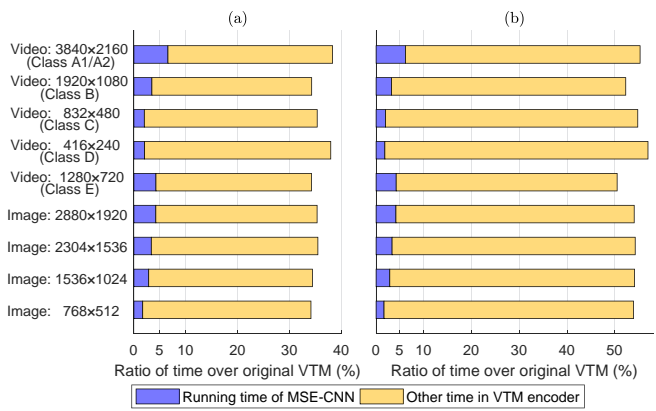}
	\caption{Running time of the proposed MSE-CNN model and the VTM encoder. (a) ``faster'' mode. (b) ``medium'' mode.}
	\label{fig:run-time}
\end{figure}

To efficiently accelerate VVC encoding, it is required that the approach itself consumes little time and space overhead. 
Thus, we analyze the running time and the space consumption of our deep MSE-CNN model, by comparing it over the original VTM 7.0 encoder \cite{VTM}.
Figure \ref{fig:run-time} shows the ratio of time for the MSE-CNN model and that for other encoding parts to overall encoding time. The results are averaged over all test sequences/images with the same resolution at four QP values.
From this figure, we can find that the time overhead introduced by MSE-CNN is less than $5\%$ for most resolutions, compared over the original VTM. For video sequences and images, the average time overhead is $3.67\%$ and $3.02\%$, respectively, which accounts for only a small part of the total encoding time. It is because the early-exit mechanism in MSE-CNN can skip most redundant checking processes in the QTMT-based CU partition. As a result, the total encoding time is averagely reduced by $64.53\%$ and $45.96\%$ at the ``faster'' and ``medium'' modes of our approach, respectively, outperforming the state-of-the-art approaches as verified in Section \ref{sec:performance}. 

With regard to space consumption, our approach also introduces little overhead. The total size of all model files are 2.9 MB, less than the 4.2 MB for the encoder of VTM anchor software. As another comparison, the input YUV-formatted video files for VTM are usually much larger than the above files, e.g., one 600-frame 1080p sequence is 1.9 GB in size. Therefore, the model files introduce almost no overhead of disk space, compared to the necessary space consumption of the VTM encoder.
In addition, Table \ref{tab:memory} tabulates the memory usage of the proposed MSE-CNN approach compared to the VTM anchor, averaged over all test video sequences with the same resolution. 
As shown in this table, the memory usage of VTM anchor ranges from 281.9 MB to 2148.4 MB, depending mainly on the frame resolution. Different from that, the MSE-CNN model holds a more stable memory usage of 157.3$\sim$180.5 MB, which changes slightly with the resolution.
Compared to the VTM anchor, the ratio of memory overhead introduced by MSE-CNN is within 55.8\% for all resolutions, and especially for high-resolution sequences (1080p or larger), that ratio decreases to 22.1\% or below. Such results have verified the memory-friendly performance of our approach.

\begin{table}
	\newcommand{\tabincell}[2]{\begin{tabular}{@{}#1@{}}#2\end{tabular}}
	\begin{center}
		\caption{Memory usage of MSE-CNN compared with the VTM anchor}
		\label{tab:memory}
		
		\begin{tabular}{|c|c|c|c|}
			
			\hline \multirow{2}{*}{Resolution} & \multicolumn{2}{c|}{Memory usage (MB)} & \multirow{2}{*}{\hspace{-0.25em}\tabincell{c}{Ratio of memory overhead\\from MSE-CNN (\%)}\hspace{-0.25em}} \\
			\cline{2-3} & MSE-CNN & VTM anchor & \\
			\hline 3840$\times$2160 & 180.5 & 2148.4 & 8.4 \\
			\hline 1920$\times$1080 & 172.5 & 780.7 & 22.1 \\
			\hline 1280$\times$720 & 166.3 & 485.8 & 34.2 \\
			\hline 832$\times$480 & 162.2 & 327.6 & 49.5 \\
			\hline 416$\times$240 & 157.3 & 281.9 & 55.8 \\
			\hline Average & 167.8 & 804.9 & 34.0 \\
			\hline
			
		\end{tabular}
	\end{center}
\end{table}

\subsection{Ablation Study}
\label{sec:ablation}

\begin{table*}[h]
	\caption{Ablation results}
	\label{tab:ablation}
	\centering
	\begin{tabular}{|c|c|c|c|c|c|c|c|c|c|}
		\hline
		\multirow{2}{*}{Ablation} & \multirow{2}{*}{Multi-stage} & \multirow{2}{*}{RD cost} & \multirow{2}{*}{Variant threshold} & BD-BR & BD-PSNR & \multicolumn{4}{|c|}{$\Delta$$T$ (\%)}\\
		\cline{7-10}
		& & & & (\%) & (dB) & QP $=22$ & QP $=27$ & QP $=32$ & QP $=37$\\
		\hline
		
		1 & & & & 6.539 & -0.299 & -59.11 & -61.56 & -62.50 & -59.34\\
		2 & \checkmark & & & 3.571 & -0.149 & -65.48 & -63.01 & -60.66 & -55.47\\
		3 & \checkmark & \checkmark & & 3.328 & -0.141 & -66.33 & -64.56 & -62.48 & -58.29\\
		4 (``faster'' mode) & \checkmark & \checkmark & \checkmark & \textbf{3.188} & \textbf{-0.134} & \textbf{-66.88} & \textbf{-65.67} & \textbf{-63.05} & \textbf{-59.57}\\ 
		\hline   
	\end{tabular}
\end{table*}

In this section, an ablation study is conducted to investigate the effectiveness of key components in our MSE-CNN approach.
Table \ref{tab:ablation} reports the ablation results averaged over all 22 test video sequences.
The ablation experiments start from a simple version of our approach, named Ablation 1, where a single-stage exit CNN (SSE-CNN), instead of MSE-CNN, is tested. In this ablation, the RD cost is not considered in the loss function of SSE-CNN ($\beta=0$ in Section \ref{sec:loss}), and the multi-threshold values are invariant to stage.
Then, multi-stage structure, RD cost and variant threshold are sequentially added to Ablation 1, named as Ablations 2, 3 and 4, respectively. Note that Ablation 4 is the ``faster'' mode of our MSE-CNN approach.
The detailed results are presented below. 
 
\textbf{Single-/multi-stage in the CNN structure:} In the SSE-CNN model, the feature maps from the same stage (Stage 2) of conditional convolution are input to all sub-networks, different from the multi-stage design of our MSE-CNN where the feature maps from various stages are used. In SSE-CNN, the number of output channels at the first layer of each residual unit is enlarged from 16 to 48, ensuring that both SSE-CNN and MSE-CNN are with the same number of trainable parameters. 
Then, the SSE-CNN model is compared with the MSE-CNN model, corresponding to Ablations 1 and 2 in Table \ref{tab:ablation}.  
As we can see, the multi-stage design achieves significantly better coding efficiency, with 2.968\% of BD-BR saving and 0.150 dB of BD-PSNR increase.

\textbf{Loss function with/without RD cost:} In the training phase of the proposed MSE-CNN model, RD cost is introduced in our loss function reflecting the coding efficiency of different split modes. Here, we compare the performance with and without RD cost in the loss function. As shown in Ablations 2 and 3 in Table \ref{tab:ablation}, the existence of RD cost can reduce the BD-BR value by 0.243\% and improve the BD-PSNR by 0.008 dB, and meanwhile the encoding time is saved by 0.85\%$\sim$2.82\% at four QP values.

\textbf{Multi-threshold variant/invariant to stage:} For implementation of the proposed MSE-CNN model, the multi-threshold values are various at different stages of CU partition, adaptive to the prediction accuracy across stages. 
To analyze its efficiency, the MSE-CNN models with both invariant and variant multi-threshold values are compared, shown as Ablations 3 and 4 in Table \ref{tab:ablation}, respectively.
In both settings, the average threshold values over all stages are 0.5 for a fair comparison. As we can see, Ablation 4 outperforms Ablation 3 by 0.140\% of BD-BR saving and 0.007 dB of BD-PSNR increase, with similar encoding time.

From the above analysis, the complexity-RD performance and complexity reduction are improved stepwise from Ablation 1 to Ablation 4. 
This verifies that all the multi-stage design, the RD cost in our loss function and the adaptive multi-threshold, are beneficial for our MSE-CNN approach. 

\begin{figure}[h]
	\centering
	\includegraphics[width=0.9\linewidth]{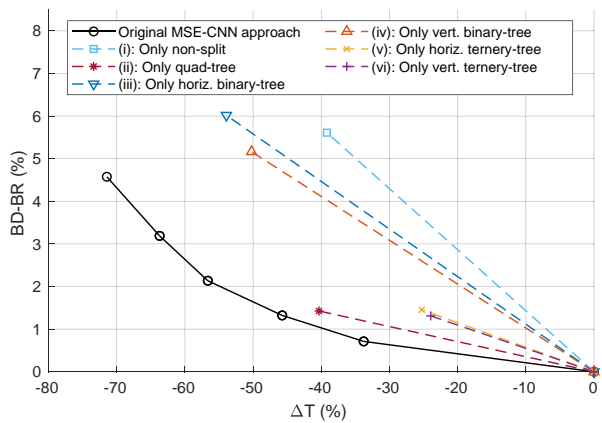}
	\caption{Complexity-RD performance for ablation settings when each individual split mode is predicted by MSE-CNN, compared with the original MSE-CNN approach. The results are averaged over all 22 test video sequences.}
	\label{fig:ablation-one-mode}
\end{figure}

In addition to the ablation study for MSE-CNN structure, it is also beneficial to analyze the settings when each individual split mode is predicted by MSE-CNN, as another ablation experiment. Figure \ref{fig:ablation-one-mode} shows the complexity-RD performance of these ablation settings, named as settings (i)$\sim$(vi). For each ablation setting, a binary classifier is used to directly decide whether to choose this mode, and the multi-threshold decision scheme is disabled.
Among all six ablation settings, setting (ii), predicting the quad-tree mode only, achieves the best in complexity-RD performance. 
It is probably because the quad-tree structure is fundamental for the CU partition and thus is the easiest to be distinguished from other modes.
In addition, the performance for two settings with symmetric split modes are close to each other, e.g., settings (iii)\&(iv) or settings (v)\&(vi) are with similar BD-BR and $\Delta T$ values.
Moreover, the complexity-RD performance for our original MSE-CNN approach is better than any ablation setting, as the original approach can always save more encoding time under similar BD-BR. 
For such performance gap, a main reason lies in the limited potential for accelerating only one split mode. As a result, even $\Delta T$ for the most time-saving ablation setting (predicting only horizontal binary-tree) fails to reach -55\%. Meanwhile, the BD-BR value for that case is up to 6\%, because of the wrongly predicted CU partition without multi-threshold decision. 
From the above analysis, it is necessary to distinguish multiple split modes for CU partition, instead of predicting only one individual mode.

\begin{figure}[h]
	\centering
	\includegraphics[width=0.85\linewidth]{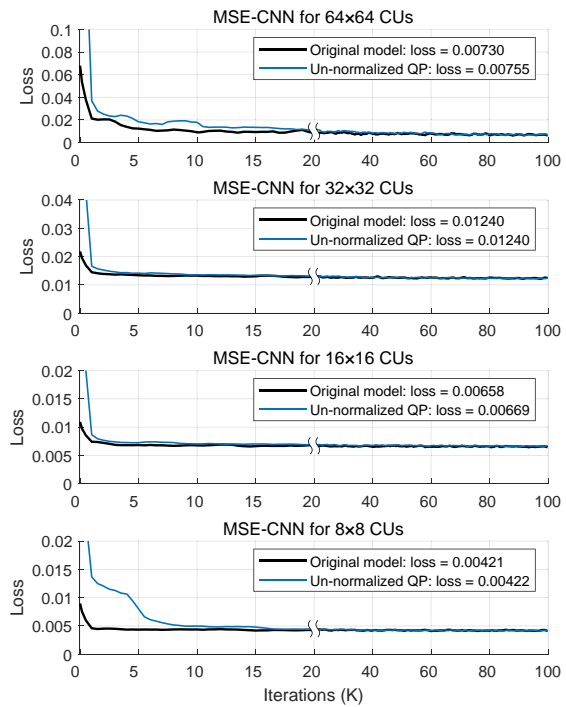}
	\caption{Loss curves for training the MSE-CNN model with different settings of QP normalization. For each curve, the converged loss value is shown in the legend.}
	\label{fig:ablation-qp}
\end{figure}

Moreover, another ablation experiment is conducted to investigate on QP normalization, containing two settings as below.
\begin{itemize}
	\item The original MSE-CNN model.
	\item Normalized QP replaced by un-normalized QP: $q$ instead of $\tilde{q}$ is introduced to MSE-CNN before the half-mask operation.
\end{itemize} 
With both settings, different MSE-CNN models can be trained correspondingly, with the loss curves for 64\ti64, 32\ti32, 16\ti16 and 8\ti8 CUs shown in Figure 10 as an example. 
As indicated in this figure, both settings can finally converge to similar loss values.
However, the initial value of loss curve for the original MSE-CNN model is smaller than that for the un-normalized setting, indicating that the original MSE-CNN model tends to converge quicker.
Thus, the effectiveness of QP normalization has been verified.

\section{Conclusion}
\label{sec:conclusion}

In this paper, we have proposed a deep learning approach to predict the QTMT-based CU partition in order to accelerate VVC encoding at intra-mode.
As VVC introduces much more flexible CU partition than HEVC, we first established a large-scale database for the diverse patterns of CU partition, and investigated the available split modes of CUs at multiple stages.
Next, we proposed a deep MSE-CNN model to determine the CU partition, combining the conditional convolution and sub-networks with sufficient network capacity.
Then, we designed an early-exit mechanism for the MSE-CNN model, which can skip the redundant checking processes on unused CUs.
Moreover, a multi-threshold decision scheme was developed, achieving a desirable trade-off between encoding complexity and RD performance.
The experimental results show that on average our approach can reduce the encoding time by 44.65\%$\sim$66.88\%, with a negligible 1.322\%$\sim$3.188\% of BD-BR increase on video sequences, outperforming other state-of-the-art approaches.

For future works, the encoding time of inter-mode VVC can also be saved with deep learning. In addition to accelerating the CU partition, there exists a potential of deep neural networks to accelerate other components in VVC, for example, intra-angular selection and motion vector estimation. Moreover, our approach may be further sped up by using various network acceleration techniques or by implementation on field programmable gate array (FPGA) devices. This can be seen as another promising future research direction for facilitating fast VVC encoders in the coming years.

\bibliographystyle{IEEEtran}
\bibliography{citeref}

\end{document}